\documentclass[journal]{IEEEtran}
\usepackage{cite}

\ifCLASSINFOpdf
  \usepackage[pdftex]{graphicx}
\else
\fi
\usepackage{siunitx}
\usepackage{amssymb}                  
\DeclareSIUnit\square{\ensuremath{\square}}
\usepackage{amsmath}
\usepackage{orcidlink}
\usepackage[font=footnotesize,justification=centering]{caption}

\ifCLASSOPTIONcompsoc
  \usepackage[caption=false,font=normalsize,labelfont=sf,textfont=sf]{subfig}
\else
  \usepackage[caption=false,font=footnotesize]{subfig}
\fi
\usepackage{comment}
\usepackage{tikz}
\usepackage[RPvoltages]{circuitikz}
\usetikzlibrary{shapes.geometric, arrows, positioning}
\usetikzlibrary{arrows.meta, bending}
\usepackage{titlesec}
\usepackage{amsfonts} 
\usepackage{amssymb}
\usepackage{enumitem,amssymb}
\usepackage{dcolumn}    
\usepackage{bm}         
\usepackage{hyperref}   
\usepackage{adjustbox}
\usepackage[justification=justified]{caption}
\usepackage[top=1.5cm, bottom=1.5cm, left=1.5cm, right=1.5cm, bindingoffset = 0pt, footskip = 0.75cm, marginparwidth = 0pt, marginparsep=0pt]{geometry}
\tikzset{addCross/.style n args={6}{
    minimum size={#5 mm}, 
    path picture={
      \draw[#6]
      (path picture bounding box.south east) -- (path picture bounding box.north west)
      (path picture bounding box.south west) -- (path picture bounding box.north east);
      \node at ($(path picture bounding box.south)!0.4!(path picture bounding box.center)$) {#1};
      \node at ($(path picture bounding box.west)!0.4!(path picture bounding box.center)$)  {#2};
      \node at ($(path picture bounding box.north)!0.4!(path picture bounding box.center)$) {#3};
      \node at ($(path picture bounding box.east)!0.4!(path picture bounding box.center)$)  {#4};
    }
  },
  addCross/.default={}{}{}{}{10}{}
}
\tikzset{mySimpleArrow/.style n args={2}{
    >={latex[#1]},
    every path/.style={draw=#2}
  },
  mySimpleArrow/.default={black}{black}
}
\tikzset{myBlockOpacity/.style n args={2}{
    every node/.style={rectangle,draw,
       minimum height=1cm,
      text=black,
      fill opacity=#1, text opacity=#2}
  },
  myBlockOpacity/.default={0.4}{1}
}
\begin{document}
%
\title{Reflection-less Filter for Superconducting Quantum Circuits}
\author{Jessica Kedziora\,\orcidlink{0000-0002-7315-5842}, \textit{Member, IEEE}, Eric Q. Bui\,\orcidlink{0009-0001-4602-2059}, Alec Yen\,\orcidlink{0000-0003-2414-7624}, Andres E. Lombo\,\orcidlink{0000-0001-5842-4314}, Kaidong Peng\,\orcidlink{0000-0003-1270-0760}, Terence J. Weir, and Kevin P. O’Brien\,\orcidlink{0000-0001-9213-6957} 
\thanks{Manuscript received xxxx. This work is based upon work supported by the Air Force Office of Scientific Research under award number FA9550-24-1-0137 and under Air Force Contract No. FA8702-15-D-0001 (Corresponding author: Jessica Kedziora je28670@mit.edu}
\thanks{{Jessica Kedziora, Eric Q. Bui, Alec Yen, Andres E. Lombo, Kaidong Peng and Kevin P. O'Brien are from the Research Laboratory of Electronics, Department of Electrical Engineering and Computer Science,  Massachusetts Institute of Technology, Cambridge, Massachusetts 02139 USA}}
\thanks{{Jessica Kedziora and Terence J. Weir are from MIT Lincoln Laboratory, Lexington, Massachusetts 02421 USA}}
\thanks{Color figures are available at: [not submitted yet]}

}

\markboth{Draft – Not Submitted}%
{Shell \MakeLowercase{\textit{et al.}}: Reflection-less filter for superconducting quantum circuits}
\maketitle{}

\begin{abstract}

Protecting superconducting quantum circuits from non-ideal return loss, including out-of-band circulator behavior and enhancing the performance of broadband quantum-limited amplifiers can be accomplished using a superconducting version of a special class of microwave filters known as reflection-less filters. These filters can simultaneously permit low pass band loss to preserve quantum efficiency and broad band reflection-less characteristics in the stop and pass bands. The filter also suppresses thermal photons emitted in its pass band from the termination resistors by the nature of the dual network topology.  This work will review the application, theory, design, and modeling of a superconducting reflection-less filter, followed by fabrication details and the presentation of cryogenic performance measurements of a monolithic device. The filter was fabricated using Al on Si, incorporating NiCr resistors, which allows for simple integration with other superconducting quantum devices. The filter with an area of 0.6 $\mathrm{\mathbf{mm^2}}$ achieves insertion loss below 1 dB, including its connectorized package over a 80\% fractional bandwidth centered at 8 GHz, and 10 dB packaged return loss from DC to above 14.5 GHz. 

\end{abstract}



\section{Introduction}
%
%
%
%

\IEEEPARstart{R}{ealization}
 of practical quantum computing requires scaling up to a large number of qubits\cite{Chow2015}.
Integration of many qubit and resonators per readout line with a few percent spacing in the resonant frequency requires careful predictability and control of coupling \cite{Li2023}, \cite{Chow2015}.  Coupling computations in design assume 50 ohm environments and mismatch, either intentional or unintentional due to inter-stage reflections changes line widths reducing predictability and limits spacing due to the spread the of the coupling rates \cite{Yen2024}, \cite{Li2023}, \cite{Sank2024}. This reduction in predictability also increases the burden of calibration during setup \cite{Sank2024}, \cite{Li2023}.

Reduction of reflections can be achieved with lossy matching but trades reduced signal to noise ratios, limiting readout speed which is also critical in scaling \cite{Ma2017}. Simply increasing the signal power is not possible because a larger amplitude causes state transition errors \cite{Bengtsson2024}. SNR can be increased by using wide-band quantum limited amplifiers such as Josephson junction based traveling wave parametric amplifiers (TWPA): however,  out of band reflection contributes to undesirable in-band gain ripple \cite{Peng2022}.  Both predictability and SNR are impacted by resonator and or qubit linewidth which is also a function of the number of thermal photons \cite{Yan2016}. The mismatch can also cause the reflection of thermal photons from one portion of a processor to another or across stages in the dilution refrigerator \cite{Ma2017}.

Stray photons are not just thermal in nature, and many qubit systems require a complex multi-signal environment. These various signals can also influence dephasing and lifetimes \cite{Yan2016}. Managing this signal environment requires filtering to route signals, protect against decay, provide multiplexing, and protect against noise sources \cite{Gao2021}, \cite{Sank2024}, \cite{Ma2017}. These filters require very low insertion loss, have high selectivity, and are normally reflective. 

The reflective nature of these filters contribute to the degradation of predictability and unintentional noise paths. This reflection is commonly mitigated with circulators and isolators instead of attenuators to mitigate the loss \cite{Gao2021}. Circulators and isolators, like reflective filters, have highly reflective out-of-band regions though, which will be shown to be especially important when using TWPA's for SNR improvement.

 Absorptive filters that maintain an impedance match in their stop band, offer an alternative to attenuators and circulators to preserve the impedance environment in the microwave signal chain in dense systems. Absorptive filters for signal lines are commonly used to limit infrared photons that use both metal powder \cite{Milliken2007} and ecosorb \cite{Gao2021}. While well suited for infrared photon control, these filters do not achieve adequate selectivity to be used in the microwave signal chain. A type of absorptive filter with good selectivity is realized using a pair of 90 degree hybrids \cite{Ma2017} where the reflection is directed to a termination, however, these are limited in bandwidth to approximately 33 \% and due to the hybrids being much larger than a reflective filter of similar bandwidth. A Lange coupler and filter in quadrature method (that can achieve up to an octave of bandwidth) was used with a pair of TWPA amplifiers in \cite{kow} using the signal-idler frequency conversion in the forward direction for gain. The reverse isolation and bandwidth are then determined by the filter absorption and stop band width. Where the output match is not at issue, this method shows how reflection-less filters may be used to potentially eliminate the output side circulator. 

Another class of absorptive filters is in \cite{morgan2011}. This design achieves low insertion loss in the pass band and good wide-band return loss.  \cite{Ge2021} implements this idea monolithically in CMOS with a variant of the topology using impedance inverters achieving greater than 6:1 reflection-less bandwidth.  Monolithic designs of the Morgan \cite{morgan2011} type are commercially available in surface mount packages from Mini-Circuits. Mini-circuits is using titanium nitride resistors which become a superconductor at 5.4 K \cite{Richardson2020} and are unsuitable for use below 1 K. The structure in \cite{morgan2011}, when assumed to operate under ideal lumped-element conditions, exhibits perfect return loss (no reflection). However, in practical implementations, factors such as delay and mutual coupling prevent it from achieving that level of performance.

\section{ Background and theory}

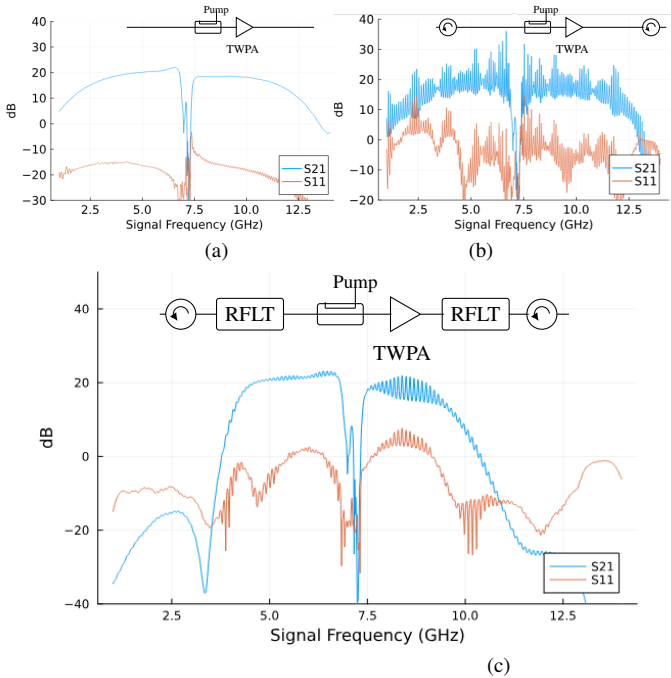
\begin{figure}[htbp]
\centering

\subfloat[]{%
\hspace*{-0.1\textwidth} 
  \begin{adjustbox}{minipage=0.45\textwidth,scale=0.45}
    
        \begin{tikzpicture}[scale=1,transform shape][very thick]
        
           
          \node[anchor=south west, inner sep=0] (img) at (0,0) {\includegraphics[width=10cm]{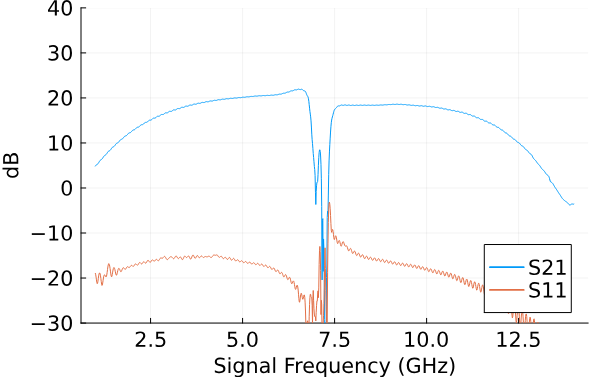}};
        
        \begin{scope}[shift={(0,13)}]
        
            \node[draw, regular polygon, regular polygon sides=3, minimum size=0.3cm, rotate=-90] (TWPA) at (7.0,-7) {};
            \node [below=0.3cm of TWPA][xshift=0.2cm, yshift=-0.1cm]{ TWPA};
        
            \begin{scope}[myBlockOpacity, rounded corners=1pt, align=center]
            \coordinate (X) at (8-2,-7);
            \node[fill=white, minimum height=10pt, text width=15pt] (CPLR) at (X) {};
            \end{scope}
            \draw[] (3.6,-7) -- (7.35-0.5,-7) node[] {};
            \draw[] (-0.5+7.75-1.5,-6.9) -- (0.5+7.75-2,-6.9) node[] {};   
            \draw[] (-0.5+7.75-1.5,-6.9) -- (-0.5+7.75-1.5,-6.8) node[] {};       
             \draw[] (0.5+7.75-2,-6.9) -- (0.5+7.75-2,-6.5) node[font=\small] {Pump}; 
        
             \draw[] (7.34,-7) -- (9.13,-7) node[] {}; 
        \end{scope}
        
        \end{tikzpicture}
  \end{adjustbox}
}
\hspace{0.02\textwidth}
\subfloat[]{%

  \begin{adjustbox}{minipage=0.45\textwidth,scale=0.45}
        \begin{tikzpicture}[scale=1,transform shape][very thick]
        
            \node[anchor=south west, inner sep=0] (img) at (0,0) {\includegraphics[width=10cm]{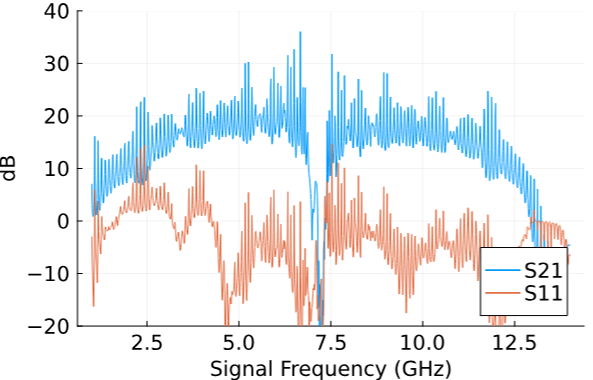}};
        \begin{scope}[shift={(0,13)}]    
            body
            \node[draw, circle, minimum size=0.5cm] (circulator) at (3.35,-7){};
             Define curved arrow with correct bending
            \draw[-{Latex[length=3,width=3]}] ([shift=(5:0.125cm)]3.35,-7) arc[start angle=5, end angle=245, radius=0.125cm];
        
            \node[draw, circle, minimum size=0.5cm] (circulator) at (9.35,-7){};
            \draw[-{Latex[length=3,width=3]}] ([shift=(5:0.125cm)]9.35,-7) arc[start angle=5, end angle=245, radius=0.125cm];
        
            \node[draw, regular polygon, regular polygon sides=3, minimum size=0.3cm, rotate=-90] (TWPA) at (7.0,-7) {};
            \node [below=0.3cm of TWPA][xshift=0.2cm, yshift=-0.1cm]{ TWPA};
        
            \begin{scope}[myBlockOpacity, rounded corners=1pt, align=center]
            \coordinate (X) at (8-2,-7);
            \node[fill=white, minimum height=10pt, text width=15pt] (CPLR) at (X) {};
            \end{scope}
            \draw[] (3.6,-7) -- (7.35-0.5,-7) node[] {};
            \draw[] (-0.5+7.75-1.5,-6.9) -- (0.5+7.75-2,-6.9) node[] {};   
            \draw[] (-0.5+7.75-1.5,-6.9) -- (-0.5+7.75-1.5,-6.8) node[] {};       
             \draw[] (0.5+7.75-2,-6.9) -- (0.5+7.75-2,-6.5) node[font=\small] {Pump}; 
        
             \draw[] (3,-7) -- (3.1,-7) node[] {}; 
        
             \draw[] (7.34,-7) -- (9.13,-7) node[] {}; 
          
           \draw[] (9.6,-7) -- (9.8,-7) node[] {};     
        \end{scope}
        \end{tikzpicture}

  \end{adjustbox}
}

\vspace{-10pt}

\subfloat[]{%
  \begin{adjustbox}{minipage=0.65\textwidth,scale=1}
        \begin{tikzpicture}[scale=0.8,transform shape][very thick]
        
          \node[anchor=south west, inner sep=0] (img) at (1,0) {\includegraphics[width=10cm]{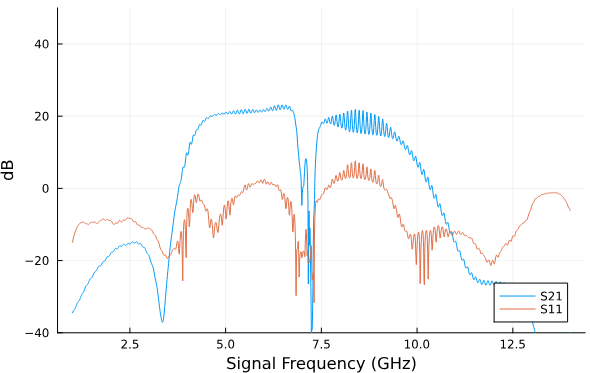}};
        
        \begin{scope}[shift={(0,12.5)}]
            \begin{scope}[myBlockOpacity, rounded corners=1pt, align=center]
            \node[fill=white, minimum height=15pt, text width=25pt] (E) at (4.5,-7){RFLT};
            \node[fill=white, minimum height=15pt, text width=25pt] (E) at (8.25,-7){RFLT};
            \end{scope}
           
            body
            \node[draw, circle, minimum size=0.5cm] (circulator) at (3.35,-7){};
             Define curved arrow with correct bending
            \draw[-{Latex[length=3,width=3]}] ([shift=(5:0.125cm)]3.35,-7) arc[start angle=5, end angle=245, radius=0.125cm];
        
            \node[draw, circle, minimum size=0.5cm] (circulator) at (9.35,-7){};
            \draw[-{Latex[length=3,width=3]}] ([shift=(5:0.125cm)]9.35,-7) arc[start angle=5, end angle=245, radius=0.125cm];
        
            \node[draw, regular polygon, regular polygon sides=3, minimum size=0.3cm, rotate=-90] (TWPA) at (7.0,-7) {};
            \node [below=0.3cm of TWPA][xshift=0.2cm, yshift=-0.1cm]{ TWPA};
        
            \begin{scope}[myBlockOpacity, rounded corners=1pt, align=center]
            \coordinate (X) at (8-2,-7);
            \node[fill=white, minimum height=10pt, text width=15pt] (CPLR) at (X) {};
            \end{scope}
            \draw[] (6.75-1.70,-7) -- (7.35-0.5,-7) node[] {};
            \draw[] (-0.5+7.75-1.5,-6.9) -- (0.5+7.75-2,-6.9) node[] {};   
            \draw[] (-0.5+7.75-1.5,-6.9) -- (-0.5+7.75-1.5,-6.8) node[] {};       
             \draw[] (0.5+7.75-2,-6.9) -- (0.5+7.75-2,-6.5) node[font=\small] {Pump}; 
        
             \draw[] (3,-7) -- (3.1,-7) node[] {}; 
             \draw[] (3.6,-7) -- (3.95,-7) node[] {}; 
             \draw[] (7.34,-7) -- (7.72,-7) node[] {}; 
             \draw[] (8.8,-7) -- (9.13,-7) node[] {};      
           \draw[] (9.6,-7) -- (9.8,-7) node[] {};     
        
           \end{scope}
            
        \end{tikzpicture}

  \end{adjustbox}
}

\caption{Comparison of TWPA performance simulated in differing configurations, each shown with the corresponding block diagram: (a) Ideal TWPA terminated with matched loads (b) Ideal TWPA terminated with measured circulators showing degraded performance with large ripple (c) TWPA terminated with non-ideal reflection-less filters (labeled RFLT) and measured circulator}
\label{fig:TWPA}
\end{figure}

We observe that quantum computing systems can benefit from low-loss, highly selective, broadband reflectionless filters. As shown in \cite{kow}, incorporating a reflection-less filter can improve reverse isolation in TWPA amplifiers. Here, we illustrate another important application: the impact of out-of-band return loss on the performance of a uniform TWPA. Using JosephsonCircuits.jl \cite{obrien} with parameters from \cite{Macklin2015} in several impedance environments a TWPA is first simulated alone with ideal terminations is shown in \ref{fig:TWPA} (a). The TWPA is then cascaded with measured isolator S-parameters, illustrating spoiling of the gain flatness in panel (b). Then, the simulation is repeated using a non-ideal reflection-less filter with the measured isolator showing substantial improvement of gain flatness in (d). This demonstrates the advantage of including the reflection-less filter with a TWPA. It provides an improved wide-band termination, and due to the low insertion loss, will have a small impact on quantum efficiency. The design, fabrication, and measurement of this filter used in the simulation is the topic of this work.

\subsection{\label{sec:level2} Synthesizing a reflection-less filter }

Consider a network described by S-parameters.  \cite{Kurokawa1965} shows where the restriction for passivity only requires $\mathbb{I}-S^\dag S \geq 0$. This implies that $1 \geq |S^\dag S| \geq 0$. For a two port network: 

\begin{multline}
\label{Sparam_cond}
|S^\dag S| = S_{11} S_{22} S^\dag_{11} S^\dag_{22} - S_{11} S_{22} S^\dag_{21} S^\dag_{12} - \\
S_{21} S_{12} S^\dag_{11} S^\dag_{22} + S_{12} S_{21} S^\dag_{12} S^\dag_{21} \geq 0   
\end{multline}

Observing that the condition can be met if both $S_{11}$ and $S_{22}$ are zero and provided 
\begin{align}
\label{Sparam_eqs}   
  1 &\geq S_{12} S_{21} S^\dag_{12} S^\dag_{21} \geq 0 \\
   1 &\geq |S_{12}|^2|S_{21}|^2 \geq 0  
\end{align}

Since there was no restriction placed on either $S_{12}$ or $S_{21}$ in frequency or phase, only that the magnitude squares must be less than one and non-negative, a frequency selective reciprocal reflection-less network is possible. The challenge is then how to find such a network for a given set of frequency selective response requirements. 

\cite{morgan2011}  showed a procedure synthesizing a reflection-less filter using lumped element reactive elements and resistors. While a similar approach to \cite{zverev1967handbook} is likely possible but difficult, Morgan demonstrates a significantly simpler approach that takes advantage of coupled mode theory \cite{Haus1987} and symmetrical network synthesis \cite{reed1956}. \cite{morgan2011} considers a symmetric two port network that can support an even and odd mode and using the approach of \cite{reed1956} and shows that for a symmetrical network with the even and odd mode reflection coefficients, $\Gamma_{even}$ and $\Gamma_{odd}$ then the two port S-parameters of the network are $S_{11} =0 $ and $S_{21} = \Gamma_{even} $ when $\Gamma_{even}=-\Gamma_{odd}$. Morgan shows that frequency selective transmission can be obtained with a filter (designed with the theory of effective parameters)\cite{zverev1967handbook} whose stop band corresponds to the pass band desired. The reciprocal network that realized the condition $\Gamma_{even}=-\Gamma_{odd}$ is then the compliment of the filter realized by exchanging each reactance with its reciprocal impedance and topology configuration \cite{zhang2019}. 

\begin{align}
\label{Sparam_Morgan_val}   
C_{lp}&=\frac{1}{L_{hp}}  & L_{lp}&=\frac{1}{C_{hp}}
\end{align}

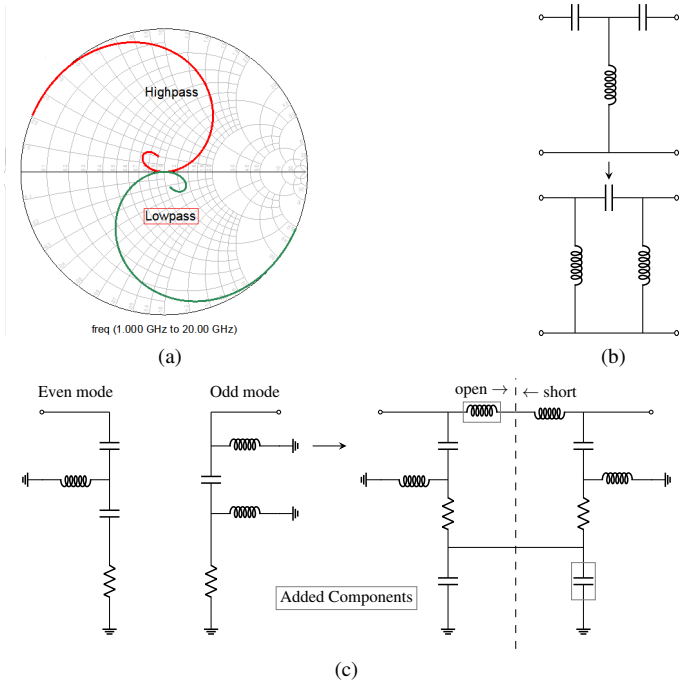
\begin{figure}[htbp]
  \centering
  \subfloat[]{%
    \includegraphics[width=0.47\linewidth]{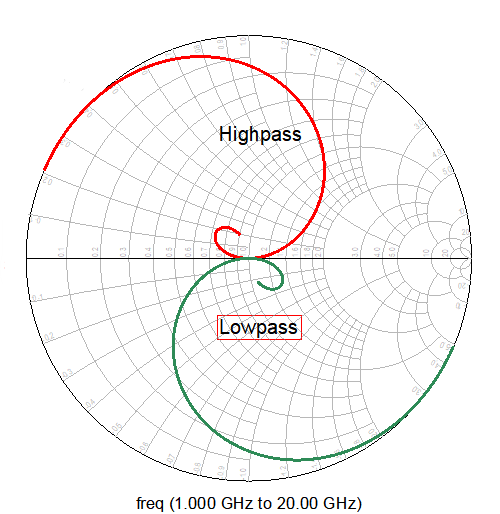}
    \label{fig:sub1}
  }
  \hfill
  \subfloat[]{%

    \begin{tikzpicture}[xscale=0.45,yscale=0.6,transform shape, line width=0.1mm]
    \ctikzset{
    	resistors/scale=0.8,
    	capacitors/scale=0.7,
    	inductors/coils=6
    }
    \draw (0,3) to[short,o-] ++(1,0);
    \draw (1,3) to[cute inductor] (1,0);
    \draw (1,3) to[C] (3,3);
    \draw (3,3) to[cute inductor] (3,0);
    \draw (0,0) to[short,o-o] (4,0);
    \draw (3,3) to[short,-o] (4,3);

    \draw [-stealth](2,3.8) -- (2,3.4);
    
    \draw (0,7) to[C] ++(2,0) to[C] ++(2,0);    
    \draw (2,7) to[cute inductor] (2,4);
    
    \draw (0,4) to[short,o-] ++(2,0) to[short,-o] ++(2,0);
    
    \draw (0,7) to[short,o-] (0,7);
    
    \draw (4,7) to[short,o-] (4,7);    
    
    \end{tikzpicture}

    \label{fig:sub2}

  }

      \vspace{-0.25cm}

     \subfloat[]{%

    \begin{tikzpicture}[xscale=0.45,yscale=0.45,transform shape, line width=0.1mm]
\ctikzset{
	resistors/scale=0.8,
	capacitors/scale=0.7,
	inductors/coils=6
}
\draw (1,6) to[short,o-] (3,6);
\draw (3,6) to[C] (3,4)
        to [C] (3,2) 
        to [R] (3,0)
        to (3,0) node[ground]{}; ;
\draw (3,4) to[cute inductor] (1,4);
\draw (1,4) -- (1,4) node[ground,rotate=-90]{}; 
\draw (8,6) to[short,o-] (6,6);
\draw (6,5) to[cute inductor] (8,5);
\draw (8,5) -- (8,5) node[ground,rotate=90]{}; 
\draw (6,6) to[C] (6,2)
        to [R] (6,0)
        to (6,0) node[ground]{}; ;
\draw (6,3) to[cute inductor] (8,3);
\draw (8,3) -- (8,3) node[ground,rotate=90]{}; 

\node[font=\Large] at (2.0,6.6) {Even mode};
\node[font=\Large] at (7,6.6) {Odd mode};

\draw [-stealth](9,5) -- (10,5);
\draw (11,6) to[short,o-] (13,6);
\draw (13,6) to[cute inductor] (15,6);

\draw (13,6) to[C] (13,4)
        to [R] (13,2) 
        to [C] (13,0)
        to (13,0) node[ground]{}; ;
\draw (13,4) to[cute inductor] (11,4);
\draw (11,4) -- (11,4) node[ground,rotate=-90]{};
\draw (19,6) to[short,o-] (17,6);
\draw (17,6) to[cute inductor] (15,6);

\draw (17,6) to[C] (17,4)
        to [R] (17,2) 
        to [C] (17,0)
        to (17,0) node[ground]{}; ;
\draw (17,4) to[cute inductor] (19,4);
\draw (19,4) -- (19,4) node[ground,rotate=90]{}; 
\draw (13,2) to[short,-] (17,2);

\draw [dashed] (15,-1) -- (15,7);

\node[font=\Large] at (14.0,6.6) {open $\rightarrow$};
\node[font=\Large] at (16,6.6) {$\leftarrow$ short};

\node[rectangle,
	draw = gray,
	minimum width = 1.1cm, 
	minimum height = 0.6cm] (r) at (14,6) {};
\node[rectangle,
	draw = gray,
	minimum width = 0.7cm, 
	minimum height = 1.1cm] (r) at (17,1) {};
\node[rectangle,
	draw = gray,
	minimum width = 1.1cm, 
	minimum height = 0.6cm,font=\Large] (r) at (10,0.5) {Added Components};
 
\end{tikzpicture}}

  \caption{Construction of reflection-less filter: (a) Reciprocal impedance relationships on smith chart, (b) dual networks and interconnection of even and odd mode networks, (c) manipulation to construct filter }
  \label{fig:FilterContruction}
\end{figure}

Fig \ref{fig:FilterContruction} (b) and (c) illustrates the generation of the reciprocal network formed by the substitute of the element and its connection in the network. Fig \ref{fig:FilterContruction} (a) illustrates the reciprocal impedance relationship on the Smith chart. At each frequency of a suitably constructed network, the summation of the network impedance and the reciprocal will be the real system impedance (center of chart), resulting in a reflection-less filter.

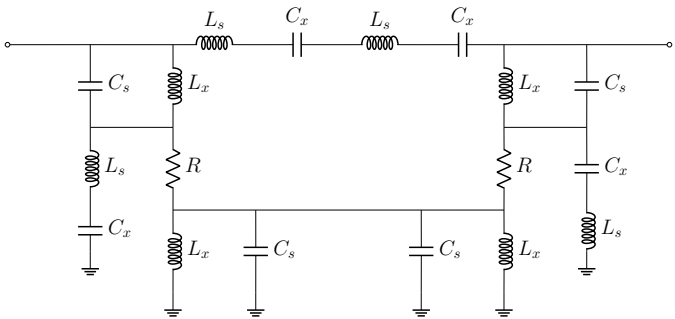
\begin{figure}

\begin{tikzpicture}[xscale=0.55,yscale=0.55,transform shape, line width=0.1mm]

\ctikzset{
	resistors/scale=0.8,
	capacitors/scale=0.7,
	inductors/coils=6
}
\draw (2,6) to[short,o-] (6,6);
\draw (18,6) to[short,o-] (14,6);
\draw (6,6) to [cute inductor, l={\Large $L_s$}] (8,6)
            to [C, l={\Large $C_x$}] (10,6)
            to [cute inductor, l={\Large $L_s$}] (12,6)
            to [C, l={\Large $C_x$}] (14,6);
\draw (4,6) to [C, l={\Large $C_s$}] (4,4);
\draw (6,6) to [cute inductor, l={\Large $L_x$}] (6,4);
\draw (14,6) to [cute inductor, l={\Large $L_x$}] (14,4);
\draw (16,6) to [C, l={\Large $C_s$}] (16,4);
\draw (6,4) to[short,-] (4,4);
\draw (14,4) to[short,-] (16,4);

\draw (4,4) to [cute inductor, l={\Large $L_s$}] (4,2)
            to [C, l={\Large $C_x$}] (4,1) node[ground]{};
\draw (16,4) to [C, l={\Large $C_x$}] (16,2)
            to [cute inductor, l={\Large $L_s$}] (16,1) node[ground]{};

\draw (6,4) to [R, l={\Large $R$}] (6,2)
            to [cute inductor, l={\Large $L_x$}] (6,0) node[ground,rotate=0]{};
\draw (14,4) to [R, l={\Large $R$}] (14,2)
            to [cute inductor, l={\Large $L_x$}] (14,0) node[ground,rotate=0]{}; 

\draw (8,2) to [C, l={\Large $C_s$}] (8,0)  node[ground,rotate=0]{}; 
\draw (12,2) to [C, l={\Large $C_s$}] (12,0)  node[ground,rotate=0]{}; 
\draw (6,2) to[short,-] (14,2);
 
\end{tikzpicture}
\caption{The band pass configuration of the reflection-less filter }
\label{fig:BPF}
\end{figure}

The band bass version of the network in \ref{fig:BPF} only requires specification of the upper and lower transmission zero frequencies, The resistors are equal to the characteristic impedance and the reactive element values are found from \cite{morgan2011} to be:
\begin{align}
\label{Morgan_BPF}  
\omega_s &=\omega_{p2}-\omega_{p1} & \quad \omega_x &=\frac{ \omega_{p2} \omega_{p1}}{ \omega_{p2}-\omega_{p1}} \\
L_s &=\frac{Z_0}{\omega_s} & \quad L_x&=\frac{Z_0}{\omega_x}\\
 C_s&=\frac{1}{Z_0 \omega_s} & \quad C_x&=\frac{1}{Z_0 \omega_x} 
\end{align}

A special feature of this filter topology that is favorable for quantum applications is that the transfer function from the resistors (when considering them as an infinite transmission line \cite{Clerk2010}) to the microwave ports is a notch filter in the filter pass band. Thermal photons generated by the filter resistors will be suppressed by the notch filter behavior in the pass band, giving a substantial advantage over an attenuator absorbing a similar amount of power. The resistors are also directionally coupled to the same side ports in the stop band. This means that the applied thermal power will radiate directionally in the filter stop band. If the quantum processor elements frequencies are in the filter pass band, the effect of heating is predicted to be reduced by several orders of magnitude. This will be shown in the heating measurement.  The transfer function from the resistor to the microwave ports is shown in \ref{fig:Heating} (f).

\subsubsection{Non-ideal behavior}

The non-ideal behavior of the filter will account for departure from being perfectly reflection-less. There are three main effects: Component and fabrication tolerance, delay primarily across the symmetry plane, and mutual coupling of the inductors. 

\subsubsection{Tolerance}

Component tolerance is the percentage difference in the actual vs. desired component values due to process variation and fabrication errors. The filter component tolerance scales together due to the filter being constructed monolithically. Scaling the inductors and capacitors together has only a small impact on the pass band. A 10\% tolerance variation uniformly degrades the return loss from perfect ($-\infty$) to approximately -20 dB. We estimate that the degradation of return loss due to tolerance is less significant than delay and mutual coupling. 

\subsubsection{Delay}
Modeling suggests delay is the largest spoiler of the reflection-less characteristic. Examining the performance of the filters below 1 GHz versus the monolithic filters in the GHz range \cite{morgan2011}\cite{lebl2019}, as well as the commercially available filters in the GHz range (minicircuits XLF-63H+), a significant difference between ideal and the achieved return loss is observed. The superior tolerance behavior of monolithic filters suggests there must be another source of return loss degradation \cite{morgan2011}\cite{lebl2019}.

\begin{figure}[htbp]
  \centering
  \subfloat[]{%
   
        \begin{tikzpicture}[xscale=0.45,yscale=0.4,transform shape, line width=0.1mm]

\ctikzset{
	resistors/scale=0.8,
	capacitors/scale=0.7,
	inductors/coils=6
}
\draw (1,4) to[short,o-] (3,4);
\draw (2,4) to [C] (2,2)
            to [R] (2,0)
            to (2,0) node[ground]{}; ;
\draw (7,4) to[short,o-] (5,4);
\draw (6,4) to [R] (6,0)
            to (6,0) node[ground]{}; ;
\draw (6,3) to[cute inductor] (8,3);
\draw (8,3) -- (8,3) node[ground,rotate=90]{}; 

\node[font=\Large] at (2.0,4.5) {Even};
\node[font=\Large] at (6,4.5) {Odd};

\draw [-stealth](9,3.5) -- (10,3.5);

\draw (11,6) to[short,o-] (13,6);
\draw (13,6) to[cute inductor] (15,6);

\draw (13,6) to [R] (13,2) 
        to [C] (13,0)
        to (13,0) node[ground]{}; ;

\draw (19,6) to[short,o-] (17,6);
\draw (17,6) to[cute inductor] (15,6);

\draw (17,6) to [R] (17,2) 
        to [C] (17,0)
        to (17,0) node[ground]{}; ;

\draw (13,2) to[short,-] (17,2);

\draw [dashed] (15,-1) -- (15,7);
\node[font=\Large] at (14.0,6.6) {open $\rightarrow$};
\node[font=\Large] at (16,6.6) {$\leftarrow$ short};

    \end{tikzpicture}
    
    \label{fig:sub1}
  }
  \vspace{-10pt}
  \subfloat[]{%

    \begin{tikzpicture}[xscale=0.4,yscale=0.4,transform shape, line width=0.1mm]

\ctikzset{
	resistors/scale=0.8,
	capacitors/scale=0.7,
	inductors/coils=6
}
\node[font=\Large] at (10,6.5) {Even};
\node[font=\Large] at (14,6.5) {Odd};

\draw (5,6) to[short,o-] (7,6);
\draw (7,6) to[cute inductor] (9,6)
            to [TL] (11,6);
\draw (7,6) to [R] (7,2) 
        to [C] (7,0)
        to (7,0) node[ground]{}; ;
\draw (7,2) to[short,-] (9,2)
            to [TL] (11,2);

\draw (19,6) to[short,o-] (17,6);
\draw (17,6) to[cute inductor] (15,6)
            to [TL] (13,6)
            to (13,6) node[ground,rotate=-90]{}; ;
\draw (17,6) to [R] (17,2) 
        to [C] (17,0)
        to (17,0) node[ground]{}; ;
\draw (17,2) to[short,-] (15,2)
             to [TL] (13,2)
             to (13,2) node[ground,rotate=-90]{}; ;


    \end{tikzpicture}
    }
    \label{fig:sub2}

\caption{The impact of delay across symmetry plane on return loss: (a) simplest element and symmetry plane connection, (b) addition of delay elements in even and odd modes to compute degraded return loss }
  \label{fig:Delay}
\end{figure}
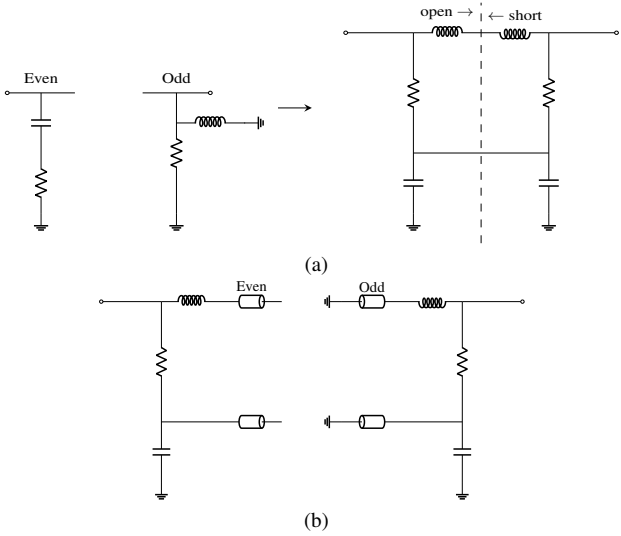

To understand the effect of delay, we start with the simplest reflection-less low-pass structure. This is a one element or just a series capacitor, which produces a shunt inductor as the dual for the odd mode. Fig \ref{fig:Delay}(a) shows the development of the single element reflection-less low-pass filter where the network transform methodology was applied. Next, delay elements are inserted on the symmetry plane to represent finite size inductors and interconnect length. These can be conveniently added to the impedance in series or parallel to find the effect as the line length grows from zero.

\noindent The component value for the low pass version:
\begin{align}
\label{Sparam_Morgan_BPF_val}   
    L = \frac{R}{\omega_p}, \quad C = \frac{1}{R \omega_p}.
\end{align}

\noindent Short- and open-circuit delay element impedance:
\begin{align}
\label{delay_elementl} 
    Z_{sc} &= j Z_l \tan(\gamma L) \\
    Z_{oc} &= -j Z_l \cot(\gamma L)
\end{align}

\noindent Odd-mode network impedance:
\begin{align}
\label{delaed_zezo} 
    Z_{1a} &= j \omega L + Z_{sc} & \quad 
    Z_{2a} &= R + \frac{\frac{-j}{\omega C} Z_{sc}}{\frac{-j}{\omega C} + Z_{sc}}, \\
    Z_{odd} &= \frac{Z_{1a} Z_{2a}}{Z_{1a} + Z_{2a}}.
\end{align}

\noindent Even-mode network impedance:
\begin{align}
\label{delaed_evenMode} 
    Z_{1b} &= R + \frac{\frac{-j}{\omega C} Z_{oc}}{\frac{-j}{\omega C} + Z_{oc}} & \quad Z_{2b} &= j \omega L + Z_{oc}, \\
    Z_{even} &= \frac{Z_{1b} Z_{2b}}{Z_{1b} + Z_{2b}}
\end{align}

\noindent Reflection coefficients:
\begin{align}
\label{delaed_oddMode} 
    \Gamma_{od} &= \frac{Z_{odd} - R}{Z_{odd} + R} & \quad\Gamma_{ed} &= \frac{Z_{even} - R}{Z_{even} + R}, \\
    \Gamma_d &= \frac{\Gamma_{od} + \Gamma_{ed}}{2}
\end{align}

Where $Z_{1a}$ is the inductor and the short circuit line; $Z_{2a}$ is the parallel combination of the resistor capacitor and short circuit line for the odd-mode circuit. $Z_{1b}$ is the parallel combination of the resistor capacitor and open-circuit line, $Z_{2b}$ is the series combination of the inductor and open-circuit line for the even-mode network. Fig \ref{fig:Delay}(b) illustrates these connections. Combining the even and odd-mode $\Gamma$ give the network return loss. When the delay ($\gamma L =0$), this collapses to the lumped network and reflection is zero. A relatively small delay in the pass-band of the filter makes a very large degradation in the return loss. The lumped approximation is usually said to be when the physical lengths are below $\frac{\lambda}{20}$. Here the lumped condition $\frac{\lambda}{20}$ is needed to be applied to the highest frequency of the stop band expected to perform which even in a monolithic structure is difficult for the large L and C values needed.

This effect can be seen in Morgan's monolithic and a printed version using the same design in \cite{lebl2019}. The absence of the problem is illustrated in the lumped element versions below 1 GHz \cite{morgan2011} where $\frac{\lambda}{20}=15$ mm at 1 GHz.

\subsubsection{Mutual coupling}

 Mutual coupling is the overlap of the magnetic field of the inductors such that the flux generated by one inductor induces a voltage on another inductor. This action also introduces cross coupling that changes the response function, because the coupling is relatively small it will be perturbative. However, this will spoil the reflection-less condition as the reflection must also be small. The mutual coupling is an all to all interaction between the inductors. Full Ansys HFSS simulations of the filter are extremely time consuming, so an alternate method to evaluate the mutual inductance search space is needed.

\begin{figure}[htbp]
\subfloat[]{%
\hspace*{-0.01\textwidth} 
  \begin{adjustbox}{minipage=0.45\textwidth,scale=0.45}
   \includegraphics[width=10.0cm]{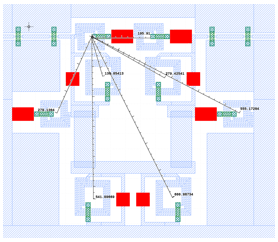}
  \end{adjustbox}
}
\hspace{0.1\textwidth}
\subfloat[]{%
\hspace*{-0.08\textwidth} 
  \begin{adjustbox}{minipage=0.45\textwidth,scale={0.5}{0.55}}
   \includegraphics[width=10.0cm]{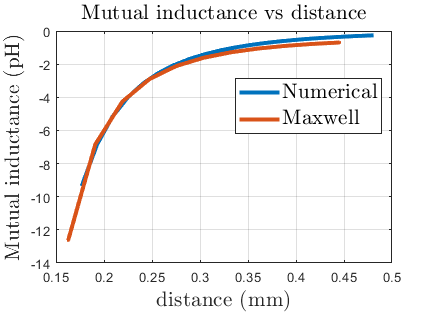}
  \end{adjustbox}
}

\vspace{-10pt}

\subfloat[]{%
\hspace*{-0.04\textwidth} 
  \begin{adjustbox}{minipage=0.43\textwidth,scale={0.5}{0.53}}
   \includegraphics[width=10.0cm]{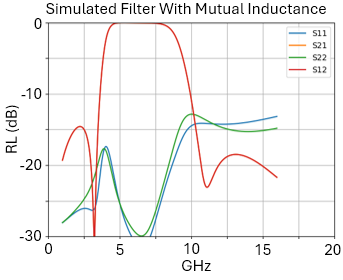}
  \end{adjustbox}
}
\hspace{0.04\textwidth}
\subfloat[]{%

  \begin{adjustbox}{minipage=0.43\textwidth,scale={0.5}{0.55}}
   \includegraphics[width=10.0cm]{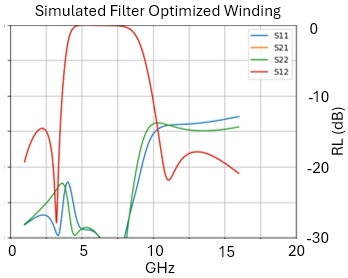}
  \end{adjustbox}
}

\caption{Illustration of the effect of mutual coupling on filter performance: (a) Path distances between inductor pairs plotted as lines on GDS, (b) mutual inductance vs distance between inductor, (c) degraded filter with delay and mutual coupling, (d) optimized filter improving in band return loss}
\label{fig:Mutual}
\end{figure}

The mutual coupling for the physical design was estimated using a numerical integration of the Neumann formula \cite{neumann1846} for a line path along the center line of the spiral inductor. This was checked with an Ansys Maxwell simulation. A script was written to generate grid point computations to produce a 2-D fit of inductor displacement in X and Y. The Neumann integral is: 
\begin{equation}
\label{Neumann_Inegral} 
   L_{m,n}=\frac{\mu_0}{4 \pi} \oint_{C_m} \oint_{Cn} \frac{d\bar x_m \cdot d \bar x_n}{|\bar x_m - \bar x_n|}
\end{equation}

where $\bar x_m$ is the vector along the differential segment. This integral is approximated by the double sum:
\begin{equation}
\label{Neumann_finite} 
L_{m,n}=\frac{\mu_0}{4 \pi} \sum_{\tilde M} \sum_{\tilde N} \frac{\Delta \bar x_m \cdot \Delta \bar x_n}{|\bar x_m - \bar x_n|}
\end{equation}

where $\Delta \bar x$ is the vector difference between the endpoints of the discretization path segment. The distances between inductors used are illustrated in Fig \ref{fig:Mutual}(a). Using this sweep, the fit allowed rapidly solving for the mutual inductance for the all to all connectivity. The magnitude of the mutual inductance vs. distance is shown in \ref{fig:Mutual} (b). The sign of the mutual inductance then comes from the winding directions in the GDS layout (Figure \ref{fig:GDS} (a)). The winding direction of the inductors is a free parameter in the design, so that optimization can be made on the orientations. There are a total of 64 possible combinations.

First, we limit the analysis to the largest couplings in order to reduce the search space down to approximately 16 cases. We then include these couplings in the ideal filter simulation by converting the mutual coupled inductors into the tee network.  The relationship between the k and the inductor values and the mutual inductance is $M=k \sqrt{L1 L2}$. $M$ and thus $k$ can be obtained from the Neumann integral as described. The series inductors are chosen to be a small value that is larger than the mutual inductance such that when the mutual inductance is subtracted, it will remain nonzero. Provided that this small value is much less than 1\% of the inductance of the inductors we are coupling, it will not have a significant impact on the circuit. Then the S-parameters of this block are computed and cascaded (using the multi-port cascade) into a composite 4 port using the S-parameters of the series tee on either side.

\begin{equation}
\label{Series Tee} 
    S_{\text{ Series tee}} = \begin{bmatrix}
        0.3333 & -0.6667 & -0.6667 \\
        -0.6667 & 0.333 & 0.6667 \\
        -0.6667 & 0.6667 & 0.3333 \\
    \end{bmatrix}
\end{equation}

\begin{figure}
\begin{circuitikz}[xscale=0.5,yscale=0.5,transform shape, line width=0.1mm]

\draw (-3,0) to[short,-] (3,0);
\draw (0,2) to[short,-] (1,2);
\draw (2,2) to[short,-] (3,2);

\draw (1,2) to[short,-] (1,6)
            to[short,-] (4,6);
\draw (2,2) to[short,-] (2,4)
            to[short,-] (4,4);

\draw (15,0) to[short,-] (9,0);
\draw (12,2) to[short,-] (11,2);
\draw (12,2) to[short,-] (11,2);

\draw (11,2) to[short,-] (11,6)
            to[short,-] (8,6);
\draw (9,2) to[short,-] (10,2)
            to[short,-] (10,4)
            to[short,-] (8,4);
            
\draw (4,6) to[cute inductor, font=\Large,  l^=$L^{'}_{1}-L_m$] (6,6);
\draw (6,6) to[cute inductor, font=\Large,  l^=$L^{'}_{2}-L_m$] (8,6);
\draw (6,6) to[cute inductor,font=\Large,   l^=$L_m$] (6,4);
\draw (4,4) to[short,-] (8,4);

\draw (-3,2) to[cute inductor, font=\large,  l^=$L_1$] (0,2);
\draw (12,2) to[cute inductor,font=\large,   l^=$L_2$] (15,2);

\node[rectangle,
	draw = gray,
	minimum width = 2cm, 
	minimum height = 3cm, font=\large] (r) at (1.5,1) {Series Tee};
\node[rectangle,
	draw = gray,
	minimum width = 2cm, 
	minimum height = 3cm, font=\large] (r) at (10.5,1) {Series Tee};
 \node[rectangle,
	draw = gray,
	minimum width = 4cm, 
	minimum height = 4cm] (r) at (6,5.5) {};

 \node[rectangle,
	draw = gray,
	minimum width = 2cm, 
	minimum height = 3cm, font=\large] (r) at (-1.5,1) {L to couple};

 \node[rectangle,
	draw = gray,
	minimum width = 2cm, 
	minimum height = 3cm, font=\large] (r) at (13.5,1) {L to couple};
 
\end{circuitikz} 
\caption{S-parameter method of including mutual inductance. Each box represents an S matrix. Series Tee's are cascaded for each mutual inductance path at each inductor. }
\label{fig:Mutual_Smethod}
\end{figure}
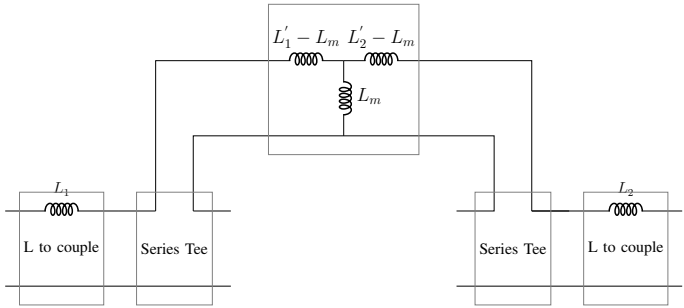

This four port can then be inserted in series with each of the inductors to be coupled. These can then be cascaded to make connections to all the other coupled inductors.  This is illustrated for one mutual connection in \ref{fig:Mutual_Smethod}.

Once the largest coupling interconnect is implemented using the computed mutual inductance matrix and the transform of the coupled inductor to the tee form, the network performance can be computed. Figure \ref{fig:Mutual} (c) shows that the 4 GHz return loss and pass-band return loss are significantly degraded by the addition of mutual coupling. By searching the permutation space of the winding direction, an improvement in Figure \ref{fig:Mutual} (d) can be achieved.

\section{Design}

\begin{figure}[htbp]
  \centering
  \subfloat[]{%
     \includegraphics[width=0.4\linewidth]{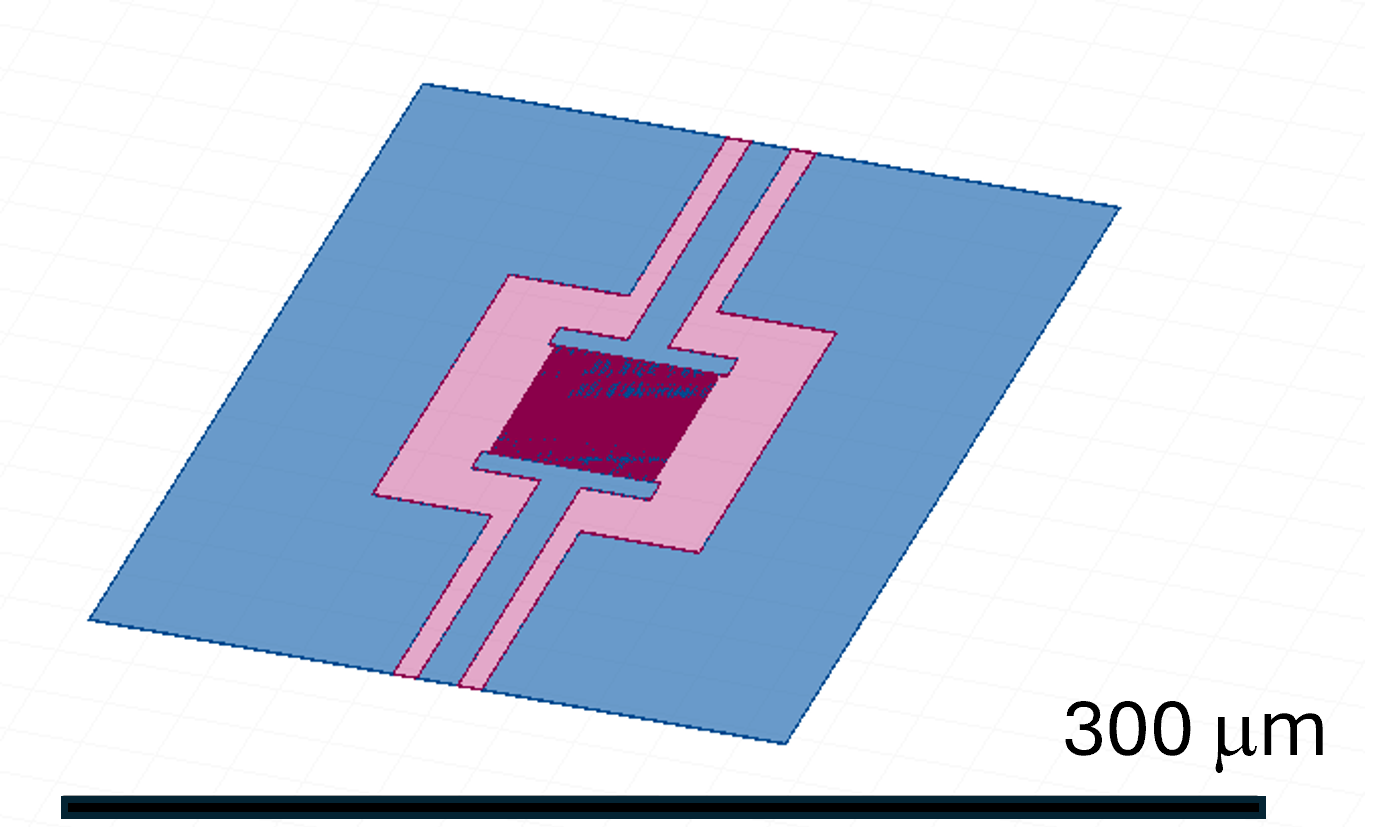}
    \label{fig:sub1}
  }
  \hfill
  \subfloat[]{%
   \includegraphics[width=0.35\linewidth]{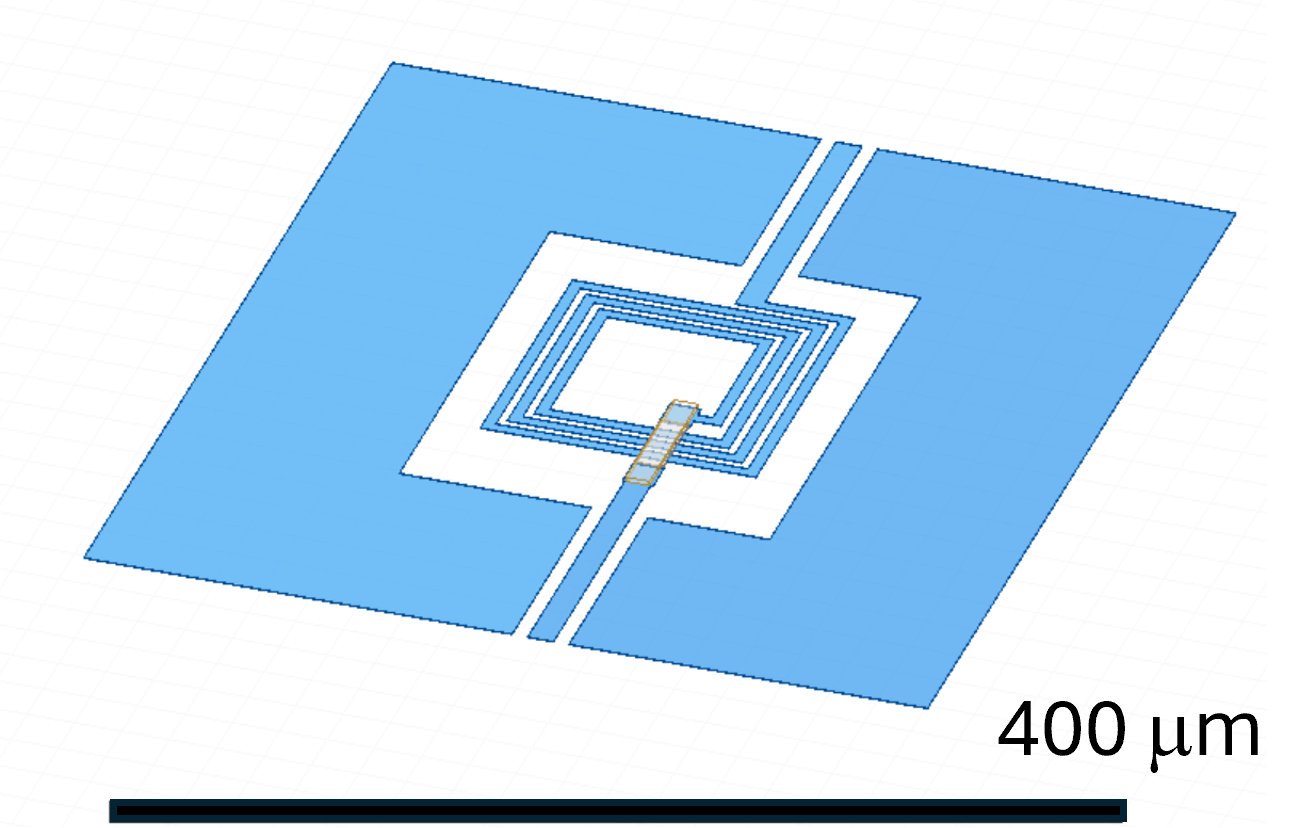}
      \label{fig:sub2}
  }

    \vspace{-0.3cm}
  \subfloat[]{%
    \includegraphics[width=0.85\linewidth]{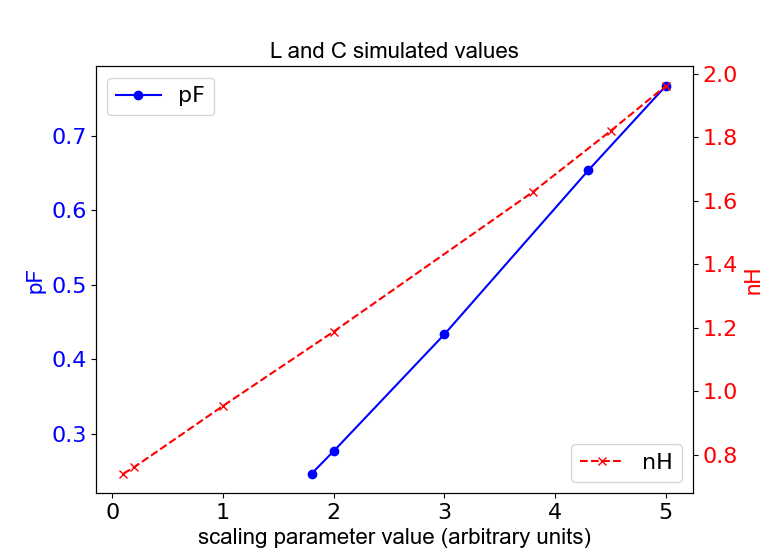}
    \label{fig:sub3}
  }

\caption{Capacitor and inductor value simulations: (a) HFSS model of capacitor, (b)  HFSS model of spiral inductor, (c) inductance and capacitance value vs a geometric scaling parameter (simulated in HFSS) with markers at the discrete values simulated. }
\label{fig:LandC}
\end{figure}

To co-fabricate the filter with other superconducting circuits, the filter design will be targeted to use the existing Al on high resistivity Si process used for qubits and TWPA's. The design begins with quantifying realizable component performance by building physical inductor and capacitor structure models according to the process capability using HFSS and Ansys Maxwell and Q3D extractor. These models are swept over a small number of discrete steps in physical parameters to build libraries of inductors and capacitors spanning the required values and tolerance. The libraries of simulated component structures permit inclusion of compensation for the non-ideal structures and delay into the S-parameters of the components. Further Python tools were written to allow fast interpolation and cascade of the S-parameters to permit use of optimization algorithms to select the final values. Figure \ref{fig:LandC} (a) shows the construction of inter-digital capacitors, and panel (b) shows the construction of the spiral inductor with an SiO2 insulted cross over. The lower panel (c)  shows the capacitance and inductance values as a function of the simulated model parameter (a unitless scaling value) with markers at the simulated values. 

The development of the resistor process requires a significant effort to obtain repeatable sheet resistance.  The selected resistor process is intended to duplicate the work of the superconducting attenuators in \cite{yeh2017}. The anticipated sheet resistance is 26 Ohms with below +/-5\% variation using 80/20 NiCr ebeam evaporated film. Currently, the repeatable sheet resistance is 40\% below the literature target value, therefore, the filter design will be modified to take into account the measured sheet resistance, at the cost of higher parasitic inductance. Less than 10\% change temperature variation from room to below 1 K was measured.

Once the component libraries are completed and the interconnect structure model is available in the S-parameter tool, the synthesis theory (eq. \ref{Morgan_BPF}) with modifications for finite delay and mutual coupling is used to obtain starting component values. The component values are interpolated using a unit-less physical parameter, allowing for automatic generation of the filter GDS. The effect of realistic interconnect structures is included by constructing GDS suitable for fabrication and simulation of the physical filter using HFSS. This finite element model of the interconnect structure has lumped ports where the inductors and capacitors will be incorporated. This generates an eighteen-port model that is cascaded with the S-parameters of the inductors, capacitors and resistors. This is optimized for response including numerically evaluated mutual coupling. After the winding directions are selected to minimize the mutual coupling effect, the physical inductor models are added to the structure, producing a new ten-port S-parameter model. This multi-port model is used for final optimization using the interpolated library capacitor models with the S-parameter cascade method. Parameter optimization was done with constrained optimization using SciPy nonlinear least squares with a frequency domain cost function and the trust region algorithm. The initial guess was found using "slide tuners" in our fast custom Python S-parameter code based on scikit-rf \cite{scikit}. 

Lastly, design of all structures to go from the on chip filter to the microwave package was accomplished by leveraging existing QCE packages and optimizing for specific geometries including development of a wide-band (1-20 GHz) wire bond launcher.  Models were constructed in HFSS and a combination of simulated and measured data was used to verify acceptable performance at the package level. A TRL calibration method to extract the connector and package S-parameters. Then, the package measured performance is cascaded with the simulated wire bond launcher and filter response to verify the packaging performance is adequate. Figure \ref{fig:chips} panel (a) inset shows the wire bond launcher with ground bridge capacitive compensation.

\begin{figure}[htbp]
\subfloat[]{%
\hspace*{-0.00\textwidth} 
  \begin{adjustbox}{minipage=0.45\textwidth,scale=0.4}
   \includegraphics[width=10.0cm]{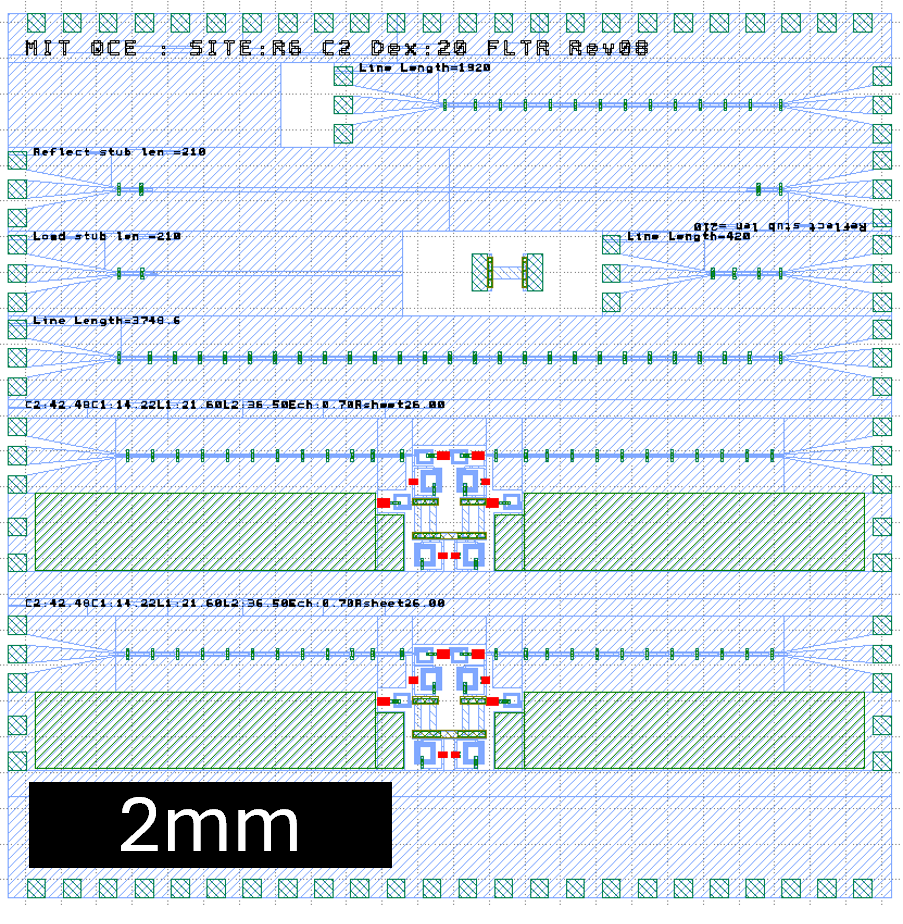}
  \end{adjustbox}
}
\hspace{0.03\textwidth}
\subfloat[]{%
  \begin{adjustbox}{minipage=0.45\textwidth,scale=0.4}
   \includegraphics[width=10.0cm]{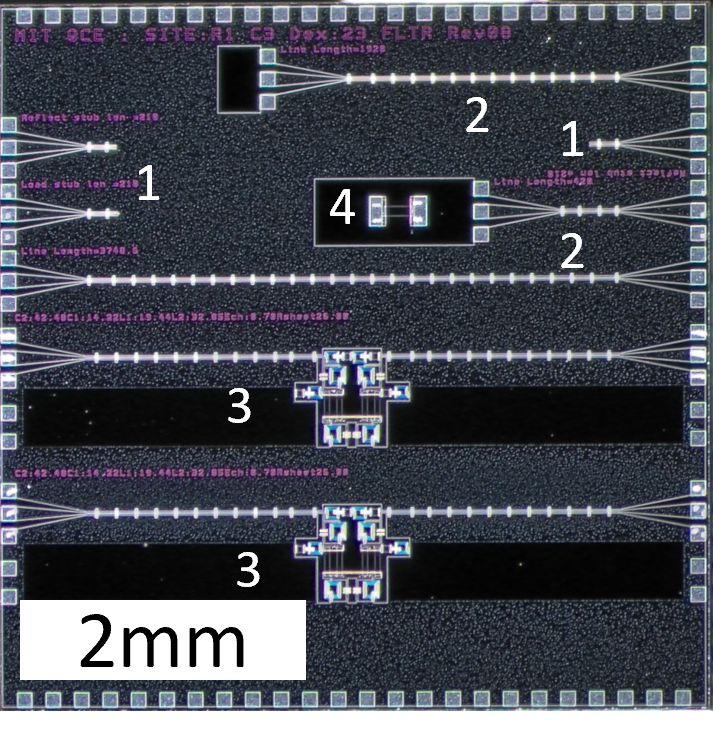}
  \end{adjustbox}
}

\caption{Final Design: (a) Final GDS for 5x5 mm chip level design including 150 $\mathrm{\mu m}$ CPW along vertical edges, cal structures as well as two filter sites. (b)  Top view of fabricated chip. Detail (1) are TRL reflect standards, (2) are TRL through and lines, (3) are the reflection-less filters  }
\label{fig:GDS}
\end{figure}

The MIT QCE GDS and additional custom Python code were used to build individual designs, assemble the chips, design variations (splits) and the wafer reticle. This step also includes developing ancillary structures with cal standards and probe pads. Figure \ref{fig:GDS} panel (a) shows the final GDS of the nominal design.

\section{Fabrication}

\begin{figure}
\centerline{\includegraphics[width=9.0cm]{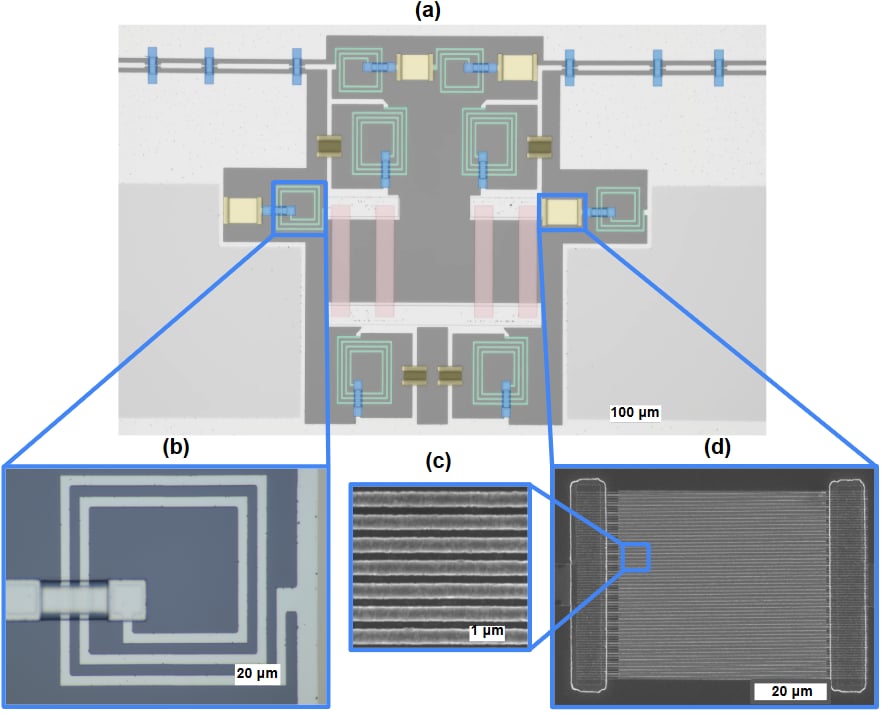}}
\caption{Fabricated devices: (a) False-color representation of the filter structure. The Nichrome resistors are highlighted in red, the finger capacitors in yellow, inductors in green, and the SiO\textsubscript{2} airbridges in blue. (b) Optical image of the spiral inductor with an SiO\textsubscript{2} airbridge (c)  Zoomed-in view of the interdigitate fingers fabricated with a width of $100\,\mathrm{nm}$ and a pitch (center‑to‑center spacing between adjacent fingers) of $200\,\mathrm{nm}$. (d) SEM image of finger capacitors, with aluminum bandages for connections. }
\label{fig:false_color}
\end{figure}

The reflection-less filters described in this paper were fabricated at MIT.nano. The final outcome of the process is depicted in Figure.~\ref{fig:false_color} and executed on a	$150\,\mathrm{mm}$ silicon wafer with an aluminum ground plane. Process modules include photolithography, reactive ion etching, electron-beam lithography, e-beam evaporation of aluminum, nichrome IV, SiO\textsubscript{2}, platinum, and titanium. Devices were fabricated on a high-resistivity ($10,000\,\Omega\!\cdot\mathrm{cm}$) [100] silicon wafer. Wafer preparation involved an SC1 organic clean using an 8:1:1 solution of deionized water, ammonium hydroxide, and hydrogen peroxide  for 15 minutes, followed by a 10 minute dip in a 16:1 volume ratio of deionized water to hydrofluoric acid solution.

Following the cleaning process, a $100\,\mathrm{nm}$ aluminum layer was deposited at a rate of $3\,\mathrm{\AA/s}$ using an e-beam evaporation system under ultra-high vacuum conditions 	($10^{-9}\,\mathrm{Torr}$). The pattern for alignment marks was defined photo-lithographically by forming a bilayer stack of PMGI SF5 and AZ3312, followed by development in AZ726 for $90\,\mathrm{s}$. A $5\,\mathrm{nm}$ titanium layer, followed by a $50\,\mathrm{nm}$ platinum layer, was then deposited using e-beam evaporation to form alignment markers, provide a conductive interface between the aluminum and nichrome resistors, and act as a heat sink for the thermal photons. Post-deposition liftoff and resist removal were carried out by submerging the sample in N-methyl-2-pyrrolidone(NMP) heated to $70\,^\circ\mathrm{C}$, followed by a descum at 25 W, with $95~\mathrm{mL/min}$ flow of O\textsubscript{2} ash in an oxygen plasma asher.

To define the ground plane, AZ3312 was spin-coated, exposed, and developed using AZ726. The aluminum layer was subsequently wet-etched at $40\,^\circ\mathrm{C}$ using Transene Aluminum Etchant Type A, followed by a spin-rinse-dry cycle. After resist removal with NMP, a $300\,\mathrm{nm}$ CSAR 62 resist layer was spin-coated, and finger capacitor patterns were exposed using an electron-beam lithography system. Next, a 100~nm Al layer was deposited at a rate of $3\,\mathrm{\AA/s}$ via the e-beam evaporation and lifted off with NMP. The resulting finger capacitors are shown in Figure.~\ref{fig:false_color} c and Figure.~\ref{fig:false_color} d highlighting the $100\,\mathrm{nm}$ wide fingers and their $200\,\mathrm{nm}$ pitch.

The SiO\textsubscript{2} bridges were patterned using a trilayer resist stack consisting of PMMA A4, PMGI SF11, and AZ3312. \cite{Dunsworth2018} The PMGI SF11 and AZ3312 layers were exposed and developed in AZ726, which contains tetramethylammonium hydroxide (TMAH). The PMMA A4 e-beam resist layer served as a protective barrier for Al against TMAH. An oxygen plasma etch was performed to remove the exposed PMMA A4. A $750\,\mathrm{nm}$ SiO\textsubscript{2} layer was then deposited as the bridge scaffold structure using an e-beam deposition system.

The same trilayer resist process was employed to pattern a $1\,\mathrm{\mu m}$ aluminum layer deposited in an e-beam evaporation system (at a rate of $3\,\mathrm{\AA/s}$) to form the bridge metal layer and to electrically connect the finger capacitors. Prior to aluminum deposition, in situ ion milling was performed to ensure proper electrical contact. Subsequently, a $75\,\mathrm{nm}$ layer of nichrome was e-beam deposited (at a rate of $1\,\mathrm{\AA/s}$) using 1/4-inch pellets in a Fabmate crucible under high vacuum conditions (approximately ($10^{-6}\,\mathrm{Torr}$)). A thickness of $75\,\mathrm{nm}$ nichrome deposition was chosen because, at this thickness, small variations have a negligible effect on the nominal \SI{25}{\ohm\per\square} sheet resistance. Furthermore, the $75\,\mathrm{nm}$ film is sufficiently conformal to climb the $100\,\mathrm{nm}$ aluminum sidewall and make reliable contact with the underlying aluminum layer.\cite{Rome1973-nichrome} 

\begin{figure} 
\centerline{\includegraphics[width=9.0cm]{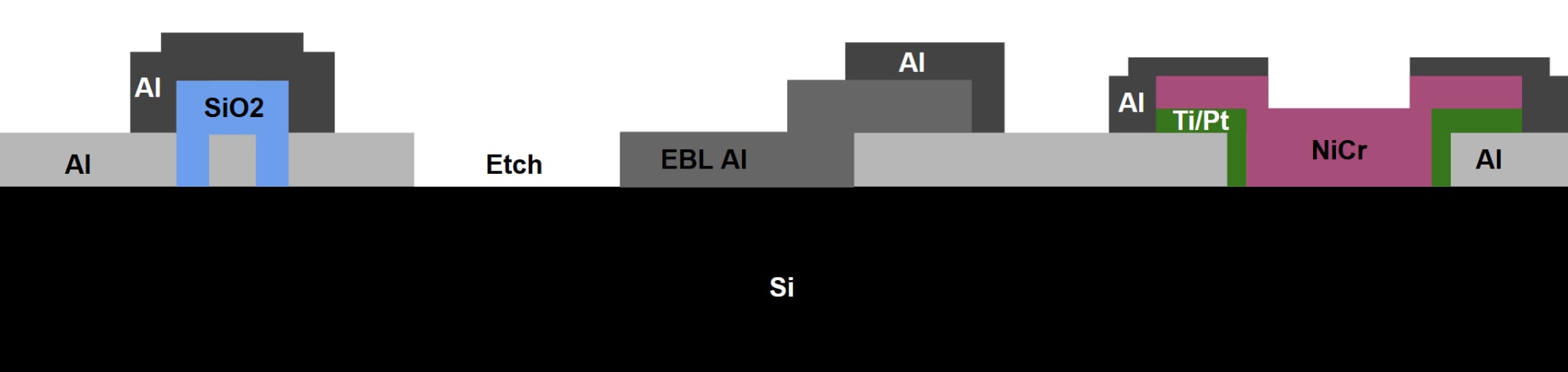}}
\caption{Cross-sectional schematic of the reflection-less filter on a high-resistivity Si substrate. The multilayer configuration comprises an Al ground plane, a SiO\textsubscript{2} layer for bridge support, EBL-patterned Al capacitors, Ti/Pt alignment markers and interconnects, NiCr resistors, and a top Al layer for air-bridge interconnects and electrical bandages.}
\label{fig:cross_section}
\end{figure}

Utilizing the same trilayer resist process described above, $600\,\mathrm{nm}$ aluminum bandages were deposited using the E-beam AJA system. Prior to deposition, low power (RF source settings of 200V, 100 mA) argon ion milling was performed to establish electrical connections between the nichrome resistors and E-beam interdigitate capacitors with the first aluminum metal layer. We fabricated the bandage structures after depositing the E-beam interdigitated capacitors, since performing ion milling beforehand would have compromised the resist features. The low-power ion milling step, together with the titanium and platinum contact layer, was employed to prevent the high contact resistance that the resistors would otherwise exhibit. The complete device stack, illustrating the multilayer structure and air-bridge formation, is depicted in the cross-sectional schematic shown in Figure.~\ref{fig:cross_section}.

\section{Measurement}

\begin{figure}
\begin{tikzpicture}[scale=0.78,transform shape][very thick]
    \draw[dashed] (0,0) -- (10,0) node[pos=0,  left] {300k};
    \draw[dashed] (0,-1.8) -- (10,-1.8) node[pos=0,  left] {4K};
    \draw[dashed] (0,-2.5) -- (10,-2.5) node[pos=0,  left] {Still};    
    \draw[dashed] (0,-3.2) -- (10,-3.2) node[pos=0,  left] {CP};    
    \draw[dashed] (0,-4) -- (10,-4) node[pos=0,  left] {MXC};     
    \begin{scope}[myBlockOpacity, rounded corners, align=center]
    \node[fill=white, minimum height=80pt, text width=1em] (E) at (3.5,-7)
    {1 \\ 2 \\ 3 \\ 4 \\ 5 \\ 6};
    \node[fill=white, minimum height=80pt, text width=1em] (E) at (6.5,-7)
    {1 \\ 2 \\ 3 \\ 4 \\ 5 \\ 6};    
    \end{scope}
    \begin{scope}[myBlockOpacity, align=center]
    \node[fill=white, minimum height=8pt, text width=2pt] (In1) at (3.1,-7){};
    \node[fill=white, minimum height=8pt, text width=2pt] (In1) at (3.9,-5.95) {};
    \node[fill=white, minimum height=8pt, text width=2pt] (In1) at (3.9,-5.95-1*0.42) {};
    \node[fill=white, minimum height=8pt, text width=2pt] (In1) at (3.9,-5.95-2*0.42)  {}   ;
    \node[fill=white, minimum height=8pt, text width=2pt] (In1) at (3.9,-5.95-3*0.42)  {} ;  
    \node[fill=white, minimum height=8pt, text width=2pt] (In1) at (3.9,-5.95-4*0.42)  {} ;
    \node[fill=white, minimum height=8pt, text width=2pt] (In1) at (3.9,-5.95-5*0.42) {};      
    \end{scope}  
    \begin{scope}[myBlockOpacity, align=center]
    \node[fill=white, minimum height=8pt, text width=2pt] (In1) at (6.9,-7){};
    \node[fill=white, minimum height=8pt, text width=2pt] (In1) at (6.1,-5.95){};
    \node[fill=white, minimum height=8pt, text width=2pt] (In1) at (6.1,-5.95-1*0.42) {};
    \node[fill=white, minimum height=8pt, text width=2pt] (In1) at (6.1,-5.95-2*0.42) {} ;   
    \node[fill=white, minimum height=8pt, text width=2pt] (In1) at (6.1,-5.95-3*0.42) {} ;   
    \node[fill=white, minimum height=8pt, text width=2pt] (In1) at (6.1,-5.95-4*0.42) {} ; 
    \node[fill=white, minimum height=8pt, text width=2pt] (In1) at (6.1,-5.95-5*0.42) {} ;     
    \end{scope}  
    \begin{scope}[myBlockOpacity, rounded corners=1pt, align=center]
    \coordinate (X) at (2,-7);
    \node[fill=white, minimum height=10pt, text width=30pt] (CPLR) at (X) {};
    \end{scope}
    \draw[] (-0.8+2,-7) -- (0.8+2+0.2,-7) node[] {};
    \draw[] (-0.5+2,-6.9) -- (0.5+2,-6.9) node[] {};   
    \draw[] (-0.5+2,-6.1) -- (-0.5+2,-5.1) node[] {};       
    \draw[] (0.5+2,-6.1) -- (0.5+2,-5.5) node[] {};  
        \draw[] (-0.5+2,-6.9) -- (-0.5+2,-6.7) node[] {};       
    \draw[] (0.5+2,-6.9) -- (0.5+2,-6.7) node[] {};  
    \begin{scope}[myBlockOpacity, rounded corners=1pt, align=center]
    \node[fill=blue, minimum height=8pt, text width=12pt, font=\small,rotate=90] (CPLATT1) at (0.5+2,-6.4){9}; 
    \node[fill=blue, minimum height=8pt, text width=12pt, font=\small,rotate=90] (CPLATT1) at (-0.8+2+0.3,-6.4){9};     
    \end{scope}

    \begin{scope}[myBlockOpacity, rounded corners=1pt, align=center]
    \node[fill=white, minimum height=8pt, text width=4pt, font=\tiny] (E) at (4.5,-5.95){O};
    \node[fill=white, minimum height=8pt, text width=15pt, font=\tiny] (E) at (4.7,-5.95-1*0.42){Term};
    \node[fill=white, minimum height=8pt, text width=4pt, font=\tiny] (E) at (4.5,-5.95-2*0.42){S};
    \node[fill=white, minimum height=8pt, text width=18pt, font=\tiny] (E) at (4.75,-5.95-3*0.42){Term2};
    \node[fill=white, minimum height=8pt, text width=15pt, font=\tiny] (E) at (4.7,-5.95-4*0.42){DUT};       
        \node[fill=blue, minimum height=8pt, text width=12pt, font=\small] (E) at (5.5,-5.95-4*0.42){20};    
    \end{scope}
    \draw[] (0.5+2,-5.5) -- (0.5+2+2.8,-5.5) node[] {};  
    \draw[] (0.5+2+2.8,-5.5) -- (0.5+2+2.8,-5.95-1*0.42) node[] {};    
    \draw[] (0.5+2+2.8,-5.95-1*0.42) -- (6,-5.95-1*0.42)  node[] {};      
    
    \draw[] (-0.5+2,-5.1) -- (-0.5+2+4.1,-5.1) node[] {};  
    \draw[] (-0.5+2+4.1,-5.1) -- (-0.5+2+4.1,-5.95-0*0.42)node[] {};      
    \draw[]  (-0.5+2+4.1,-5.95-0*0.42) -- (6,-5.95-0*0.42) node[] {};    
    \draw[]  (4,-5.95-0*0.42) -- (4.35,-5.95-0*0.42) node[] {};
    \draw[]  (4,-5.95-1*0.42) -- (4.35,-5.95-1*0.42) node[] {};
    \draw[]  (4,-5.95-2*0.42) -- (4.35,-5.95-2*0.42) node[] {};
    \draw[]  (4,-5.95-3*0.42) -- (4.35,-5.95-3*0.42) node[] {};
    \draw[]  (4,-5.95-4*0.42) -- (4.35,-5.95-4*0.42) node[] {};
    \draw[]  (4,-5.95-5*0.42) -- (6,-5.95-5*0.42) node[] {};
    \draw[]  (5.05,-5.95-4*0.42) -- (5.2,-5.95-4*0.42) node[] {};
    \draw[]  (5.8,-5.95-4*0.42) -- (6,-5.95-4*0.42) node[] {};
    \draw[]  (-0.8+2,-4) -- (-0.8+2,-7) node[] {};
    \draw[]  (-0.8+2,-3.45) -- (-0.8+2,-1.75) node[] {};  
    \draw[]  (-0.8+2,-1.2) -- (-0.8+2,1) node[] {};   
    \begin{scope}[myBlockOpacity, rounded corners=1pt, align=center]
    \node[fill=blue, minimum height=4pt, text width=10pt, font=\small, rotate=90] (IA1) at (-0.8+2,-3.72){10};
    \node[fill=blue, minimum height=4pt, text width=10pt, font=\small, rotate=90] (IA2) at (-0.8+2,-1.5){10};  
    \end{scope}
    \begin{scope}[myBlockOpacity, rounded corners=1pt, align=center]
    \node[fill=blue, minimum height=4pt, text width=10pt, font=\small, rotate=90] (OA1) at (8,-1.5){6};
    \end{scope}
    \draw[]  (7,-7) -- (8,-7) node[] {};
    \draw[]  (8,-7) -- (8,-1.75) node[] {};
    \draw[]  (8,-0.50) -- (8,1) node[] {};
    \draw[]  (8,-1.2) -- (8,-0.94) node[] {};
    \node[draw, regular polygon, regular polygon sides=3, minimum size=0.3cm] (LNA) at (8,-0.8) {};
    \node [right=0.1cm of LNA]{Cryo LNA};
    \node[draw, regular polygon, regular polygon sides=3, minimum size=0.3cm,rotate=90] (RT1) at (7,1) {};
    \node[draw, regular polygon, regular polygon sides=3, minimum size=0.3cm,rotate=90] (RT2) at (6,1) {};
    \node[above right=0.1cm of RT1]{A1};
    \node[above right=0.2cm of RT2]{A2};
    \begin{scope}[myBlockOpacity, rounded corners=3pt, align=center]
    \node[fill=white, minimum height=20pt, text width=40pt, font=\small] (VNA) at (4.0,1){VNA};
    \end{scope}
    \draw[]  (8,1) -- (7.15,1) node[] {};
    \draw[]  (6.7,1) -- (6.15,1) node[] {};
    \draw[]  (5.7,1) -- (4.79,1) node[] {};  
    \draw[]  (-0.8+2,1) -- (3.2,1) node[] {};

\end{tikzpicture}

\caption{Dilution refrigerator calibrated S\textsubscript{11} and normalized through measurement system. Lower stages are labeled in accordance with  BlueFors documentation: Still, CP "cold plate" and MXC is "mixing chamber". Vertical number boxes are multi-pole switches. Shaded boxes represent attenuators where numbers are dB. Input attenuators were selected to trade minimizing room temperature noise with dynamic range. The reflectometer coupler (on MXC) is a 20 dB coupler. The cryogenic low noise amplifier (LNA) has 36 dB gain. The room temp gain is 34 dB. The cal standards on the switch are labeled "O" open, "Term" termination, "S" short, "Term2" for a 6 dB return loss reflection. The 20 dB attenuator after the DUT permits S\textsubscript{11} measurement and normalized through. A1 and A2 are room temperature amplifiers. VNA is a vector network analyzer instrument. Not shown is an offset short also used in a second cool down.  DC control of the switches is not shown.  }
\label{fig:FridgeS11}
\end{figure}
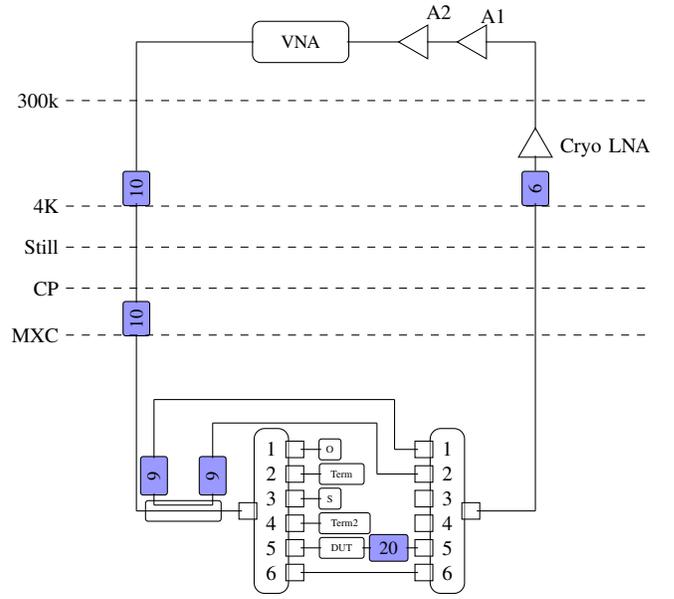

The filter is measured at 20 mK in a BlueFors cryogenic helium dilution refrigerator. The actual S\textsubscript{11} and normalized (to a through line) S21 characteristics are needed to prove the design and fabrication were successful. Calibrated return loss is difficult due to the inaccessibility of the connections and the large amount of required loss to bring the room temperature thermal noise low enough to prevent heating. Furthermore, the many interconnects cascaded presents a far from 50 ohm environment. \cite{Wang2021-hi} Describes a method using a circulator and a switched microwave path to perform a vector S\textsubscript{11} calibration below 100 mK. This work uses a similar switched standards approach with a different configuration to achieve a much broader bandwidth and provides simultaneous normalized through capability. The measurement system is installed in the refrigerator shown in figure \ref{fig:FridgeS11}. The input line attenuators are selected to provide adequate dynamic range and limiting room temperature noise. The output line attenuator before the 4 K LNA is selected to improve its broad band return loss without reducing dynamic range unacceptably. The input line attenuation to the filter device is ~45 dB at 8 GHz. The output path gain is -5 dB including all loss at 8 GHz, where the 4 K LNA has 36 dB gain, the remainder of the 32 dB gain being external at room temperature. The gain slope of the entire path is approximately -5 dB per GHz. Measurements are made as S21 using a network analyzer. The S21 measurement ratios for the reflectometer and calibration are performed in Python post-processing.  The left-hand switch selects which load is connected to the coupler. The coupler arms are selected by the right-hand switch for forward or reflected samples or the through paths. As the right-hand switch is a reflecting type, 9 dB attenuators are added in the coupler arms. The coupler directivity is limited by this method to 18 dB and this will be seen to be adequate post-correction. 

The measurement sequence is to connect the standards and the filter (DUT) in turn (corresponding to positions 1-5 on the left switch) and sequence the coupler in the FWD and REV coupled arms using the right hand switch (positions 1 and 2). After the standards and filter are measured, the through paths are selected (DUT) corresponding to position 5 and 6.

The measurement of the filter was completed with two cool-downs of the dilution refrigerator with 3, 4 and 5 calibration standards. When 5 standards were used, the 5th was in the position of the through line. The standards were characterized at room temperature such that a data-based calibration can be applied by post-processing the collected data. Calibration is based on a 3-term error model \cite{rytting}. Calibration is performed by using least squares at each frequency point of 
\begin{equation}
\label{cal_Solution} 
     Solve: \rightarrow \begin{bmatrix}
        1 & \Gamma_1 \Gamma_{m1}  &  \Gamma_1\\
        1 & \Gamma_2 \Gamma_{m2}  & \Gamma_2\\
        1 & \Gamma_3 \Gamma_{m3}  & \Gamma_3\\   
        \vdots & \vdots & \vdots 
    \end{bmatrix} 
         \begin{bmatrix}
         e_{00}   \\
        e_{11}   \\
         \Delta_e    \\         
    \end{bmatrix}      
     =
         \begin{bmatrix}
         \Gamma_{m1}   \\
        \Gamma_{m2}   \\
         \Gamma_{m3}   \\        
         \vdots
    \end{bmatrix}  
\end{equation}

Then applying the error terms using 
\begin{equation}
\label{cal_apply} 
\Gamma_{corrected}=\frac{\Gamma_m - e_{00}}{\Gamma_m e_{11} - \Delta_e}
\end{equation}

\subsection{ Cryogenic measurement challenges}

It is observed from the error gain expression and \cite{Rehnmark1974} that the S\textsubscript{11} calibration will perform poorly if the standards or the reference plane change significantly due to matching ( how similar the S-parameters are of the differing switch paths) and tracking (how the changes are relative to each other over temperature) of the switch paths. To quantify this, a Monty Carlo S-parameter simulation was performed with first guesses of the matching and tracking errors based on room temperature and 77 K cable characterizations, then refined with corrected-standards data. This simulation varied the noise according to the anticipated vector network analyzer (VNA) uncertainty in the 10 Hz measurement bandwidth and power ($\pm$ 0.4 dB), and applied both phase errors ($\pm$ 10°) and variation of the return loss magnitude (nominal $\pm$ 10 dB), and insertion loss (nominal $\pm$ 0.5 dB) of the separate paths.  Here, nominal means the measured room temperature characteristics. Figure \ref{fig:calerror} (a) shows a gray-scale heat map of how the measured return loss over the 500 trials of the Monty Carlo simulation. Darker areas indicate more likely measurement values. The standards will change at cryogenic temperatures due to the contraction and mismatch of the material's temperature coefficients.

To observe how the tracking and matching error degrades performance, a notional error locus for the standards uncertainty is plotted in blue in Figure \ref{fig:calerror} (c); then mapping that over the Monty Carlo tracking an matching error produces the red constellations. Because there are many wavelengths the error covers the min and max standing wave ratio (SWR) phase rapidly over frequency. This will mean that the true data should be equal to or less than the calibrated data with error. To quantify the error and determine what return loss can be measured with confidence, the error in the corrected standards at 20 mK is compared to the room reference data shown in Figure \ref{fig:calerror} (b). The standards' error exceeds 0.316 (10 dB) at 13 GHz. As a result, the measured return loss of 10 dB or better should be observable up to 13 GHz.

The measurement system included cables between the switch and the standards and DUT. This was to place the reference plane at the DUT attached to the experiment cold plate. These cables will degrade the calibration performance due to tracking and matching errors between paths. However, this is weighted against the effect of having an uncalibrated cable before the DUT with unknown return loss and impedance at 20 mK. This trade is made by comparing the cal degradation with the degradation due to the uncalibrated cable. \cite{Glasser1978} developed an expression for the increase in error due to uncertainty or "error gain" derived from the cascading of 2 port networks and the fact that this cascading is a bilinear transform mapping circles to circles.   

\begin{align}
    \label{error_gain} 
    \delta_{2M}&=\left | \frac{(1-\Gamma_L S_{22})^2}{S_{11}S{22}-\Delta} \right | \delta_{1M}\\
    \Delta &=S_{11}S_{22}-S{12}S_{21}
\end{align}

Here $\delta_{1M} $ is the error locus before the additional network and $\delta_{2M}$ is the total error locus after the addition of the error network. Thus the quantity in the absolute value acts as an error gain. The measured return loss of the cables used at room temp and 77 K is worse than 16 dB above 11 GHz.  The filter anticipated packaged return loss is on the same order magnitude. As there are several wavelengths in cable (not including the cables to avoid additional tracking errors), the observable 10 dB return loss is limited to occur at or below 11 GHz.  The measured residual calibration error outperforms this, but not substantially. As a result, the measured data includes the cables to move the reference plane to the DUT and include the larger than the switch alone tracking and matching errors.

\section{ Measured data}
\begin{figure}[htbp]

\subfloat[]{%
\hspace*{-0.02\textwidth} 
  \begin{adjustbox}{minipage=0.9\textwidth,scale=0.8}
   \includegraphics[width=10.0cm]{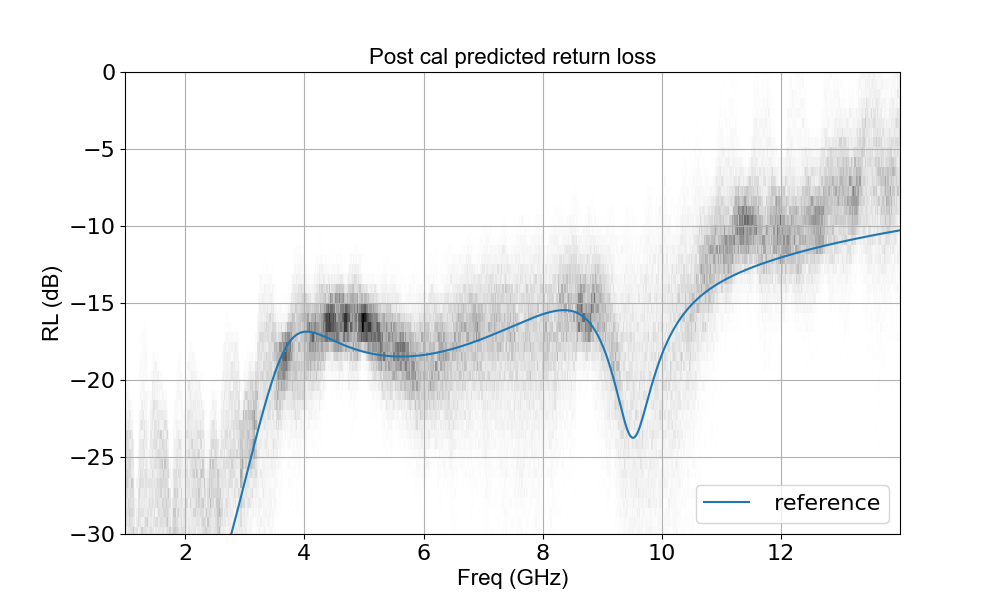}
  \end{adjustbox}
}
\vspace{-20pt}
\subfloat[]{%
\hspace*{-0.00\textwidth} 
  \begin{adjustbox}{minipage=0.5\textwidth,scale=0.5}
   \includegraphics[width=10.0cm]{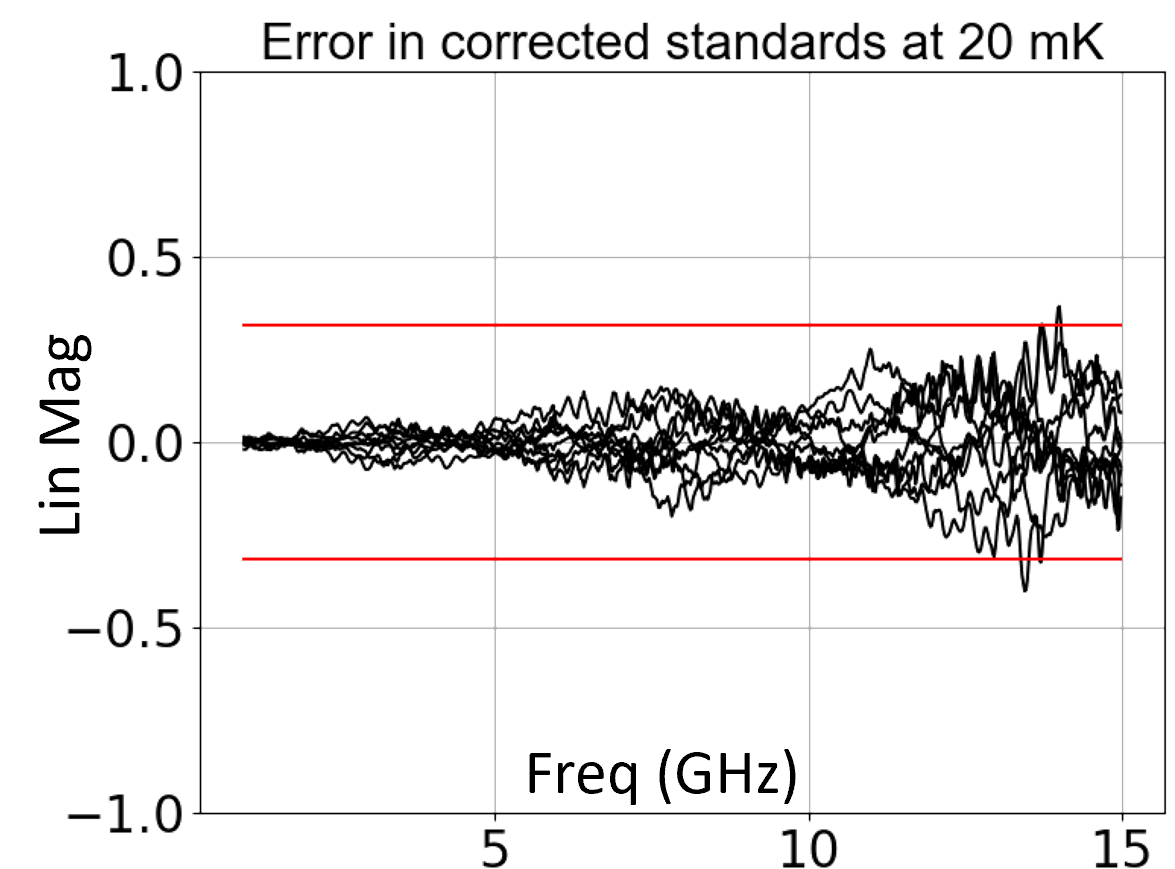}
  \end{adjustbox}
}
\hspace{0.0\textwidth}
\subfloat[]{%

  \begin{adjustbox}{minipage=0.4\textwidth,scale=0.35}
 \includegraphics[width=10.0cm]{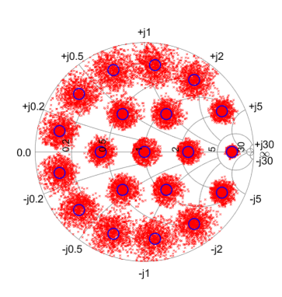}
  \end{adjustbox}
}

\caption{System cal simulation: (a) Monty Carlo of corrected S\textsubscript{11} with all errors. trace is expected return loss, gray distribution is range of possible measured return loss with tracking and matching errors in dilution refrigerator. (b) Comparison of corrected standards to stored measured data, showing with horizontal lines where measurement error will exceed 10 dB return loss. (c) Estimated error locus of standards room to cold in dark circles, where associated point constellations showing growth due to tracking and matching errors)}
\label{fig:calerror}
\end{figure}

\begin{figure}
\centerline{\includegraphics[width=10.0cm]{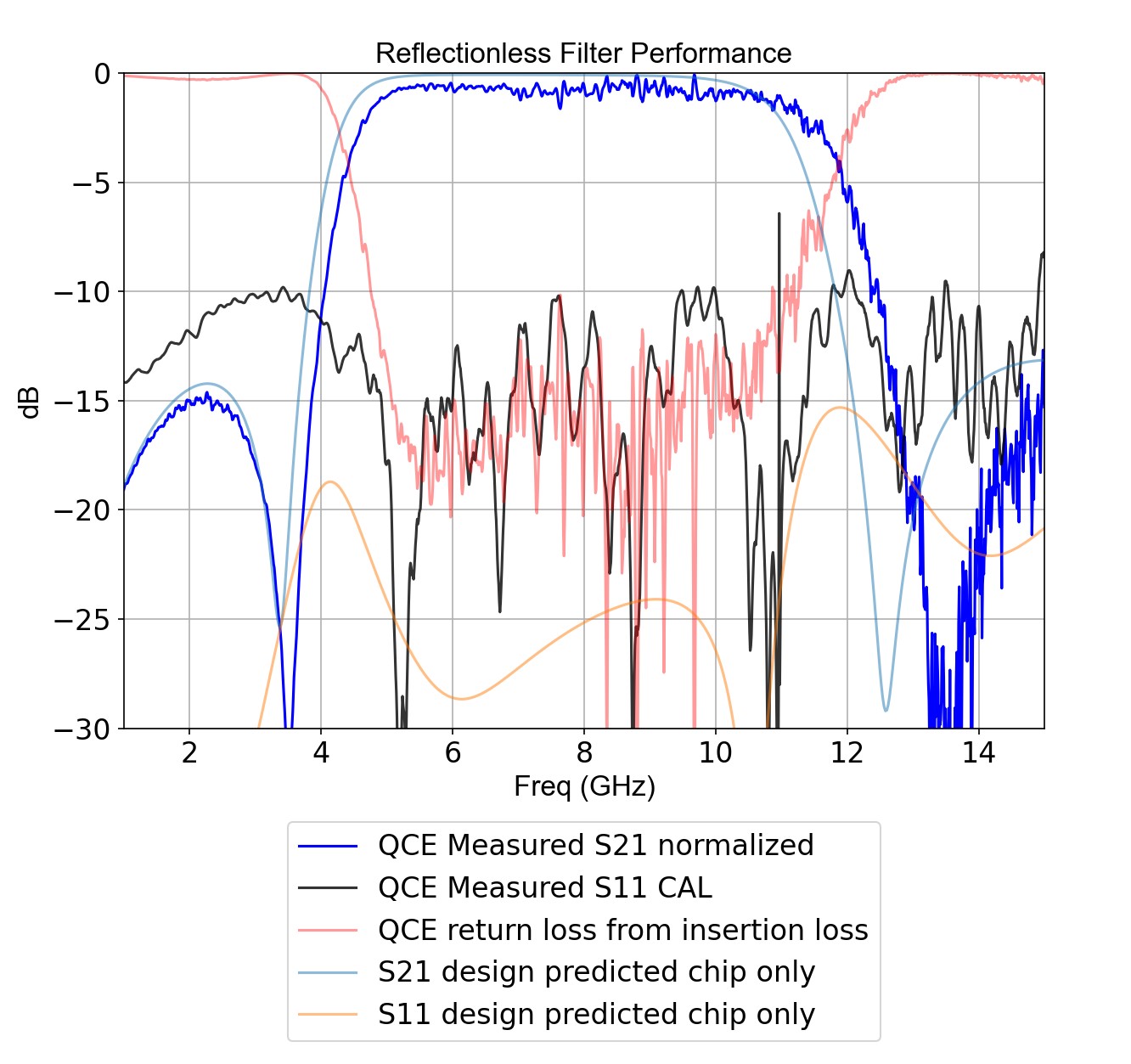}}
\caption{ (a) Measured performance of the superconducting reflection-less filter device in a connectorized package at 20 mK with design data for comparison 
}
\label{fig:measuredData}
\end{figure}

\begin{figure}[htbp]
\subfloat[]{%
\hspace*{-0.02\textwidth} 
  \begin{adjustbox}{minipage=0.3\textwidth,scale=0.25}
   \includegraphics[width=10.0cm]{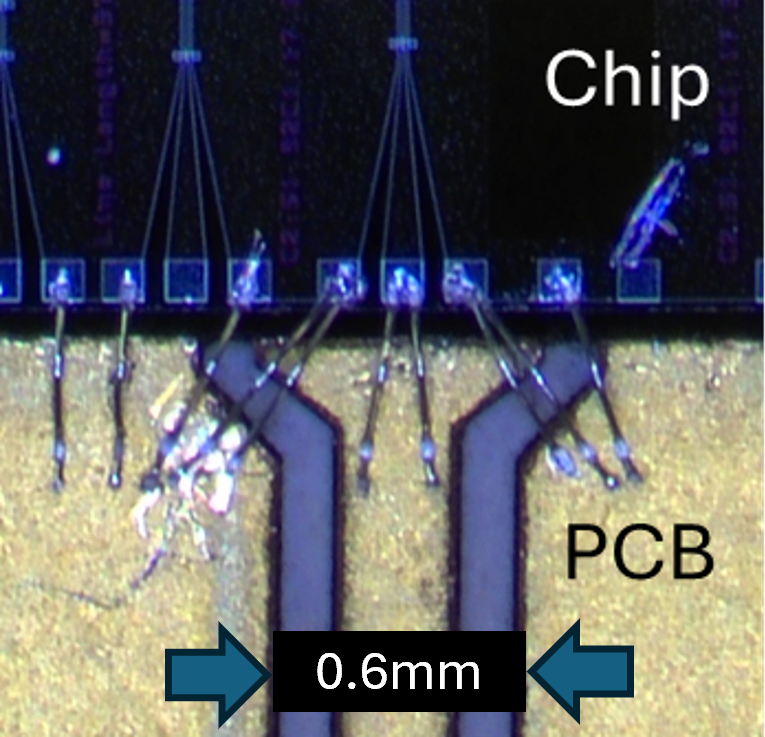}
  \end{adjustbox}
}
\hspace{0.07\textwidth}
\subfloat[]{%

  \begin{adjustbox}{minipage=0.3\textwidth,scale=0.3}
   \includegraphics[width=10.0cm]{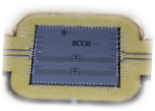}
  \end{adjustbox}
}
\hspace{0.07\textwidth}
\subfloat[]{%
\hspace*{-0.00\textwidth} 
  \begin{adjustbox}{minipage=0.3\textwidth,scale=0.3}
    \includegraphics[width=10.0cm]{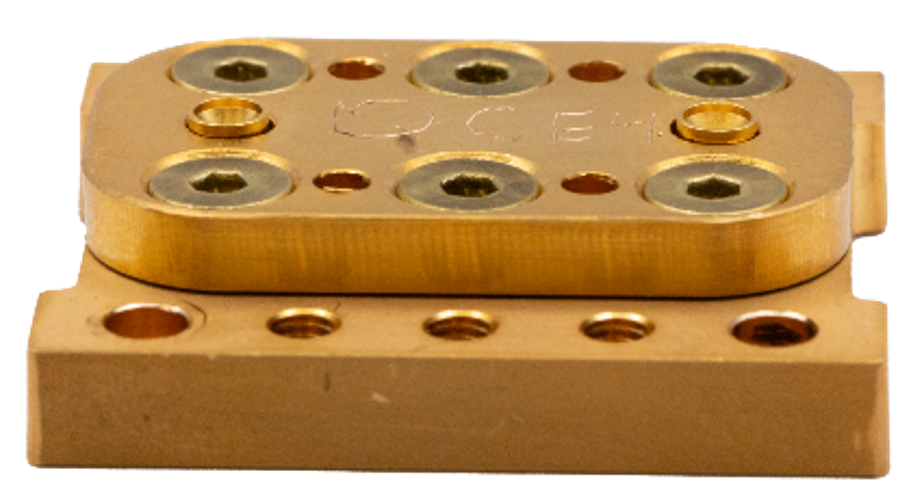}
  \end{adjustbox}
}

\caption{ (a) Detail of wire bond launch showing winged line adding capacitance to compensate for wire bond inductance. (b) Shows the chip wire  bonded onto the TMM10i CPW carrier circuit board. (c) Shows the microwave package ready to install in the dilution refrigerator. 
}
\label{fig:chips}
\end{figure}

  An image of measured  die on the CPW carrier and the microwave package is illustrated in Figure \ref{fig:chips} (b) and (c). The measured performance of the connectorized device at 20 mK is shown in Figure \ref{fig:measuredData}. The return loss from the passivity limit of the normalized through line is also plotted. It overlays with the calibrated S\textsubscript{11} data up to 9 GHz giving confidence to the S\textsubscript{11} data. Comparing the loss heat map (Figure \ref{fig:calerror} (a)) and the measured data, the measured packaged data is very similar in the pass band region and is worse in the lower stop band. The filter that was measured had resistors with low sheet resistance. The design target was 26 Ohms, and the realized value was 16 Ohms. The incorrect sheet resistance limits the stop-band return loss to 13.5 dB. Table \ref{tab:comparison_reflection-less} compares this filter to similar monolithic devices in literature. The device achieves similar performance to the enumerated examples with much lower insertion loss due to its superconducting nature and cryogenic operation. There is an anomalous spike in the calibrated S\textsubscript{11} data. As the measurement power is very high, it is unlikely to be a spurious two-level system.  There is no corresponding S21 dip so it is believed this is related to a setup resonance. 

\begin{table}[!t]
\caption{COMPARISON WITH OTHER REFLECTION-LESS BANDPASS FILTERS}
\label{tab:comparison_reflection-less}
\footnotesize
\resizebox{\columnwidth}{!}{%
  \begin{tabular}{|p{1.2cm}|p{1.8cm}|p{1.3cm}|p{1.2cm}|p{1.2cm}|p{1.9cm}|p{1.5cm}|}
    \hline
    \textbf{\centering Source} &
    \textbf{\centering \begin{tabular}[c]{@{}c@{}}Center\\Frequency\\ \& 3 dB BW\end{tabular}} &
    \textbf{\centering Insertion Loss (dB)} &
    \textbf{Return Loss (dB)} &
    \textbf{\centering \begin{tabular}[c]{@{}c@{}}Broad-\\band\\Return\\ Loss(dB)\end{tabular}} &
    \textbf{\centering \begin{tabular}[c]{@{}c@{}}Stopband\\Attenuation(dB) 
    \end{tabular}} &
    \textbf{\centering \begin{tabular}[c]{@{}c@{}} Area \\ $\lambda_0 \times \lambda_0$ \end{tabular}}

    \\ \hline

        \begin{tabular}[c]{@{}l@{}}%
    This Work \\ (Packaged)
    \end{tabular} & 
    \begin{tabular}[c]{@{}c@{}}%
    8\,GHz Center \\ 6.4\,GHz BW%
    \end{tabular} & 0.35 & 10 & 10 & 15 & 0.02 $\times$ 0.02 \\ \hline

    \begin{tabular}[c]{@{}l@{}}%
    \cite{morgan2011}
    \end{tabular} &
    \begin{tabular}[c]{@{}c@{}}%
    200\,MHz Center \\ 100\,MHz BW%
    \end{tabular} & $\sim$2.6 & $\sim$26 & 12 & 43 & -  \\ \hline

    \begin{tabular}[c]{@{}l@{}}%
    \cite{Li2021} (Chip)
    \end{tabular} & 
    \begin{tabular}[c]{@{}c@{}}%
    3.9\,GHz Center \\ 1.2\,GHz BW%
    \end{tabular} & 3 & 15 & 10 & 19 & 0.01 $\times$ 0.02 \\ \hline
    \begin{tabular}[c]{@{}l@{}}%
    \cite{Simpson2021} (Chip)
    \end{tabular} & 
    \begin{tabular}[c]{@{}c@{}}%
    9.7\,GHz Center \\ 1.84\,GHz BW%
    \end{tabular} & 6.2 & 11 & 11 & 18 & 0.01 $\times$ 0.03 \\ \hline

    \begin{tabular}[c]{@{}l@{}}%
    \cite{Liu2020}\\ (Simulated)
    \end{tabular} & 
    \begin{tabular}[c]{@{}c@{}}%
    3.78\,GHz Center \\ 1.11\,GHz BW%
    \end{tabular} & 1.33 & 30 & 11 & 12& -  \\ \hline


        \begin{tabular}[c]{@{}l@{}}%
    \cite{minicircuitsReflectionlessPass} \\ (Packaged)
    \end{tabular} & 
    \begin{tabular}[c]{@{}c@{}}%
     6 GHz cutoff \\ (low pass)%
    \end{tabular} & 1.3--1.6 & $\sim$ 17 & 5  & 12--15 & -  \\ \hline
  \end{tabular}%

}
\end{table}

\section{ Stop band heating}

Controlling excess thermal photons in quantum computing environments is critical. \cite{Simbierowicz2024}\cite{yeh2017}  Heating and excess thermal photons impact the linewidth and lifetime of qubits, which in 
 turn impact the ability to correct errors.\cite{Simbierowicz2024}\cite{Yan2016} Since the filter is absorptive in its stop band, any power from a control, pump and readout signals may heat the filter.\cite{Gao2021}  Pump signals for four wave parametric amplifiers are in the range of -80 to -23 dBm \cite{Guarcello2024} \cite{Faramarzi24}. Squid based traveling wave parametric amplifiers pump power is up to -47 dBm \cite{Gaydamachenko25}. These large signals may be in the pass or stop band and as a result the impact of heating by these signals must be examined. The readout and attenuated control signals are small enough so that no significant heating is expected. (Microwave powers used in readout are -120 to -90 dBm \cite{Gunyh2024}. Qubit control signals, once attenuated at the readout line are typically in the same power range \cite{Sah2024}.)

\begin{figure}[htbp]

\subfloat[]{%
\hspace*{-0.03\textwidth} 
  \begin{adjustbox}{minipage=0.35\textwidth,scale=0.45}
   \includegraphics[width=10.0cm]{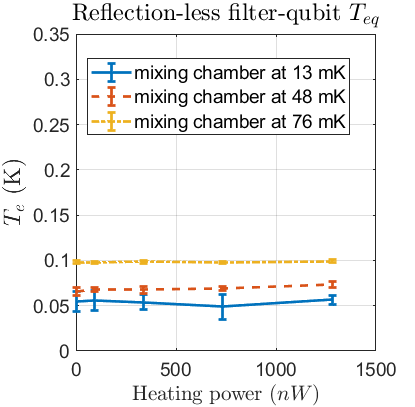}
  \end{adjustbox}
}
\hspace{0.06\textwidth}
\subfloat[]{%

  \begin{adjustbox}{minipage=0.55\textwidth,scale=0.55}
    \includegraphics[width=10.0cm]{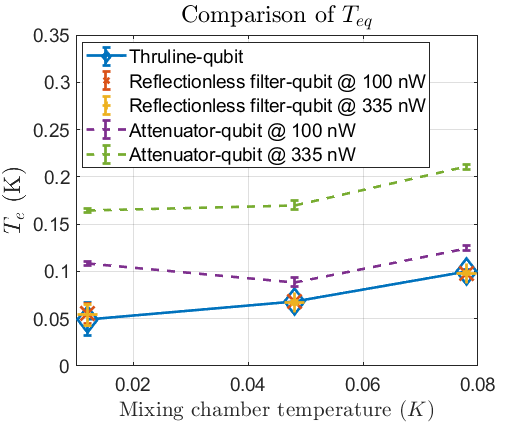}
  \end{adjustbox}
}

\vspace{-10pt}

\subfloat[]{%
\hspace*{-0.00\textwidth} 
  \begin{adjustbox}{minipage=0.65\textwidth,scale=0.7}
        \begin{tikzpicture}[scale=1,transform shape][very thick]
        
        \begin{scope}
            \begin{scope}[myBlockOpacity, rounded corners=1pt, align=center]
            
            \node[fill=white, minimum height=15pt, text width=25pt] (biasT) at (3.5,-7){biasT};     
            \node[fill=white, minimum height=15pt, text width=25pt] (RFLT) at (5,-7){RFLT};
            \node[fill=white, minimum height=15pt, text width=25pt] (DCB) at (6.5,-7){DCB};     
            \node[fill=white, minimum height=15pt, text width=25pt] (qubit) at (8,-7){qubit};

            \end{scope}

           \node (in) at (1,-7){Drive and Readout};
           \node[font=\small] (DC) at (3.5,-6.3){DC};
           \node (out) at (11,-7){Readout};
           
            body
       
            \node[draw, circle, minimum size=0.5cm] (circulator) at (9.35,-7){};
            \draw[-{Latex[length=3,width=3]}] ([shift=(5:0.125cm)]9.35,-7) arc[start angle=5, end angle=245, radius=0.125cm];

           \end{scope}

            \draw[->] (in) -- (biasT);
            \draw[] (biasT) -- (RFLT);
            \draw[] (RFLT) -- (DCB);
            \draw[] (DCB) -- (qubit);
            \draw[] (qubit) -- (circulator);
             \draw[->] (circulator)--(out);      
             \draw[->] (DC)--(biasT);    
        \end{tikzpicture}
  \end{adjustbox}
}

\vspace{-10pt}
\subfloat[]{%
\hspace*{-0.00\textwidth} 
  \begin{adjustbox}{minipage=0.65\textwidth,scale=0.7}
        \begin{tikzpicture}[scale=1,transform shape][very thick]
        
        \begin{scope}
            \begin{scope}[myBlockOpacity, rounded corners=1pt, align=center]
            
            \node[fill=white, minimum height=15pt, text width=25pt] (biasT) at (3.5,-7){biasT};     
            \node[fill=white, minimum height=15pt, text width=25pt] (ATTN) at (5,-7){ATTN};
            \node[fill=white, minimum height=15pt, text width=25pt] (DCB) at (6.5,-7){DCB};     
            \node[fill=white, minimum height=15pt, text width=25pt] (qubit) at (8,-7){qubit};

            \end{scope}

           \node (in) at (1,-7){Drive and Readout};
           \node[font=\small] (DC) at (3.5,-6.3){DC};
           \node (out) at (11,-7){Readout};
           
            body
       
            \node[draw, circle, minimum size=0.5cm] (circulator) at (9.35,-7){};
            \draw[-{Latex[length=3,width=3]}] ([shift=(5:0.125cm)]9.35,-7) arc[start angle=5, end angle=245, radius=0.125cm];

           \end{scope}

            \draw[->] (in) -- (biasT);
            \draw[] (biasT) -- (ATTN);
            \draw[] (ATTN) -- (DCB);
            \draw[] (DCB) -- (qubit);
            \draw[] (qubit) -- (circulator);
             \draw[->] (circulator)--(out);      
             \draw[->] (DC)--(biasT);    
        \end{tikzpicture}
  \end{adjustbox}
}
\vspace{-10pt}

\subfloat[]{%
\hspace*{-0.00\textwidth} 
  \begin{adjustbox}{minipage=0.65\textwidth,scale=0.7}
        \begin{tikzpicture}[scale=1,transform shape][very thick]
        
        \begin{scope}
            \begin{scope}[myBlockOpacity, rounded corners=1pt, align=center]
               
            \node[fill=white, minimum height=15pt, text width=25pt] (qubit) at (8,-7){qubit};

            \end{scope}

           \node (in) at (1,-7){Drive and Readout};
           \node (out) at (11,-7){Readout};
           
            body
       
            \node[draw, circle, minimum size=0.5cm] (circulator) at (9.35,-7){};
            \draw[-{Latex[length=3,width=3]}] ([shift=(5:0.125cm)]9.35,-7) arc[start angle=5, end angle=245, radius=0.125cm];

           \end{scope}

            \draw[->] (in) -- (qubit);
            \draw[] (qubit) -- (circulator);
             \draw[->] (circulator)--(out);      
  
        \end{tikzpicture}
  \end{adjustbox}
}

\vspace{0pt}
\vspace{-20pt}

\subfloat[]{%
\hspace*{-0.00\textwidth} 
  \begin{adjustbox}{minipage=0.65\textwidth,scale=0.85}
    \includegraphics[width=10.0cm]{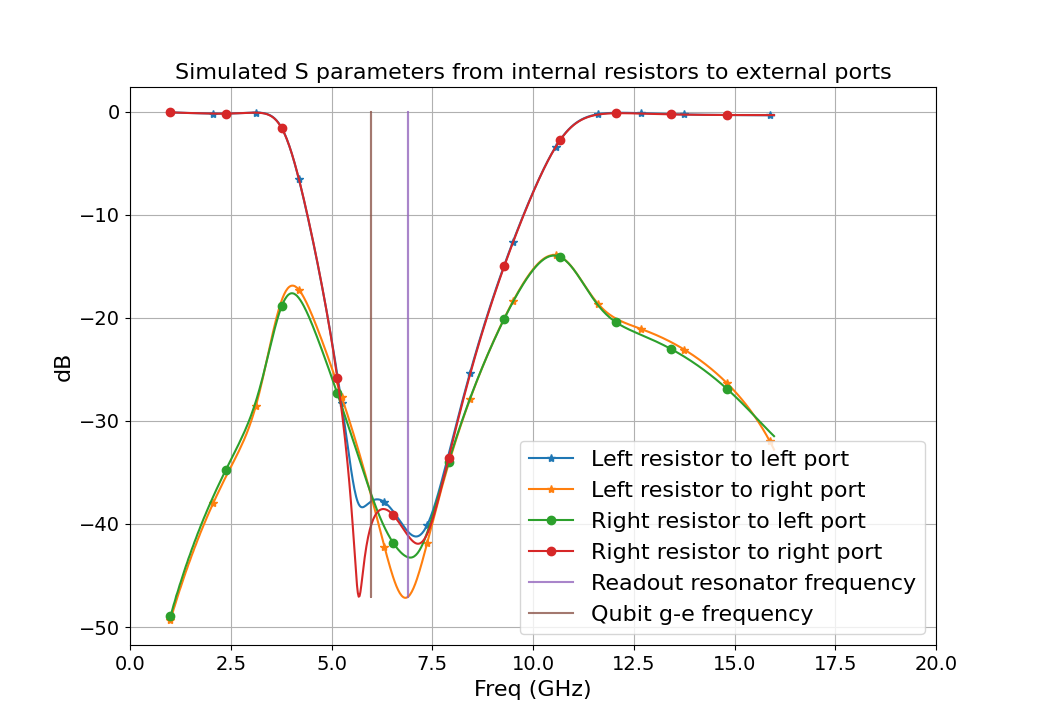}
  \end{adjustbox}

  }
  \caption{(a) Effective temperature of the qubit device computed using measured T\textsubscript{1} and T\textsubscript{2E} with the method in \cite{Yan2016} at three mixing chamber temperatures with the reflection-less filter swept over the DC heating power. DC heating power is dissipated in the filter resistors. The effective temperature is independent of the heating power showing the thermal photon suppression. (b) Comparison of the reference configuration (thru line) qubit effective temperature as a function of the mixing chamber temperature with the reflection-less filter and attenuator configurations. The reflection-less filter with heating power applied overlays on the reference thru line configuration for both applied heating powers. The attenuator configuration shows substantial effective temperature increase under the same heating power. (c) Block diagram of qubit device based heating measurement setup. The reflection-less filter resistors are heated with a DC bias. The reflection-less filter is followed with a high pass filter acting as a DC block. (d) The same setup using a resistive cryogenic attenuator instead of the reflection-less filter for comparison. (e) Thru line with the qubit device for a baseline reference measurement.  Not shown in (c), (d) and (e) are fridge cabling, 4 K LNA and room temp electronics. (f) Simulated filter response to power injected at resistors showing nulling of thermal photon power at RF ports. }
  \label{fig:Heating}
\end{figure}
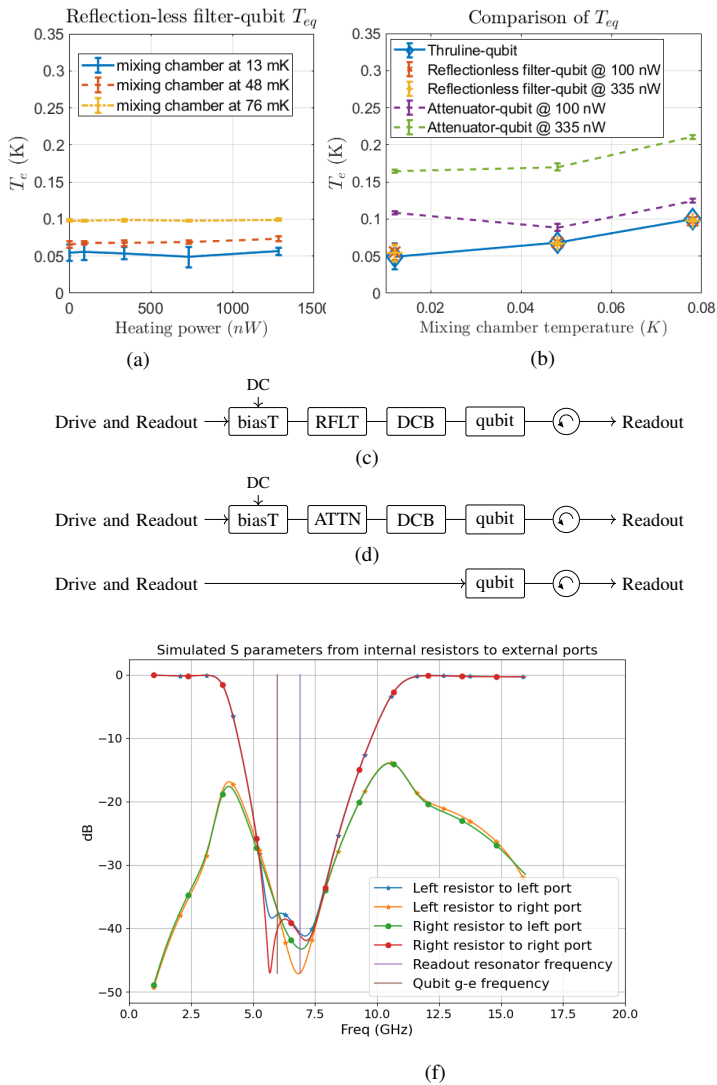

To investigate the effect of heating of the filter's resistors by large signals, the filter is used with an inline transmon qubit coupled to a readout resonator (qubit device). This is compared to the same configuration substituting a thru line (as a reference) and a cryogenic resistive attenuator in the reflection-less filter position. This is shown in figure \ref{fig:Heating} (c),(d) and (e). The readout resonator is measured to have a frequency of $\omega_r/2\pi=6.894$~GHz with linewidth of $\kappa/2\pi=0.28$~MHz. The transmon qubit is measured to have a frequency of $\omega_{q}/2\pi=5.974$~GHz, anharmonicity of $\alpha/2\pi=-199$~MHz, and total dispersive shift of $2\chi/2\pi=-0.33$~MHz. As the filter has a substantial suppression of the noise power spectral density in the pass band where the qubit and resonator frequencies are, and the DC block acts as a filter below and above the filter pass band, no effect should be observed. The proximity of the qubit and resonator frequencies to the features of the plot in \ref{fig:Heating}(f) is a coincidence. 

The filter or attenuator resistors are heated by applying a DC current calculated with the room-temperature resistance of the filter or attenuator resistance. The measured resistor change is less than 2\% down to 4 K. The DC current is derived from a battery and resistive divider routed via DC wiring and a bias tee on the mixing chamber. A high pass filter was used as a DC block after the filter to prevent DC from reaching the qubit shown in \ref{fig:Heating} (c) and (d). The base temperature was swept using the mixing chamber heater, allowing for at least one hour between reaching the temperature set point and making a measurement, similar to \cite{yeh2017}. DC power was swept from near zero up to 1.2 $\mathrm{\mu W}$. 1.2 $\mathrm{\mu W}$ is sufficient to cause observable heating of the mixing chamber and was used as a diagnostic to verify the filter was being heated.

Figure \ref{fig:Heating} (a) shows the effective temperature of the qubit device with the DC heated reflection-less filter in line.  The energy decay rate (T\textsubscript{1}) and the dephasing rate with Hahn echo (T\textsubscript{2e}) of the qubit \cite{Krantz2019} were measured and the effective temperature was solved for using the model in \cite{Yan2016}. For the reflection-less filter, the effective temperature at three mixing chamber temperatures is independent of the DC heating power. Figure \ref{fig:Heating} (b) shows the behavior over the mixing chamber temperature set point for the three configurations: Filter, attenuator, and through line reference. The reflection-less-filter configuration overlays with the thru-line configuration showing the application of 100 and 335 nW causes no increase in the qubit device's effective temperature. The attenuator configuration shows a substantial rise in the effective temperature over the same two heating powers. This characteristic gives the reflection-less filter substantial utility over an attenuator, as the latter, when reducing a signal’s amplitude, injects broadband thermal photons.

\section{Conclusion}
The reflection-less band pass filter provides an elegant solution to return-loss induced  problems in quantum processors such as resonator linewidth changes and gain ripple in traveling-wave parametric amplifiers.  Our filter demonstrated low insertion loss and more than 14:1 reflection-less bandwidth. The filter also was shown to have a valuable suppression of thermal photons from large signal heating in its pass band due to the construction from dual networks.  We showed degradation in the theoretical in band and high frequency return loss and other filters in literature is due to symmetry-plane delay and mutual coupling effects. These effects were incorporated into the design using an S-parameter-based methodology where only portions of the filter required full wave FEM simulation - a method that dramatically reduces the design time.  The filter was fabricated using a qubit compatible process with cryogenically suitable NiCr resistors added at the last step. The filter size of less than 0.6 $\mathrm{mm^2}$ is readily compatible for co-fabrication on other TWPA and circuit QED devices. The filter design and fabrication were successful as shown by the vector calibrated S\textsubscript{11} below 10 dB from 1 to 14 GHz and low loss via normalized through measurements operating below 20 mK. The filters measured cryogenic performance was similar to other monolithic non-cryogenic designs with lower insertion loss due to the cryogenic superconducting nature. Finally, the heating suppression feature was verified experimentally by characterizing its performance with a qubit coupled to a resonator. The demonstrated performance makes the filter suitable for integration into circuit QED based quantum processors.

\section*{Acknowledgment}
Jessica Kedziora would like to thank the Lincoln Scholars Program at MIT Lincoln Laboratory, and especially Katrina Silwa and Mollie E. Schwartz for their support and guidance. 

\bibliographystyle{IEEEtran}
\bibliography{IEEEabrv,references}

\begin{thebibliography}{10}
\providecommand{\url}[1]{#1}
\csname url@samestyle\endcsname
\providecommand{\newblock}{\relax}
\providecommand{\bibinfo}[2]{#2}
\providecommand{\BIBentrySTDinterwordspacing}{\spaceskip=0pt\relax}
\providecommand{\BIBentryALTinterwordstretchfactor}{4}
\providecommand{\BIBentryALTinterwordspacing}{\spaceskip=\fontdimen2\font plus
\BIBentryALTinterwordstretchfactor\fontdimen3\font minus \fontdimen4\font\relax}
\providecommand{\BIBforeignlanguage}[2]{{%
\expandafter\ifx\csname l@#1\endcsname\relax
\typeout{** WARNING: IEEEtran.bst: No hyphenation pattern has been}%
\typeout{** loaded for the language `#1'. Using the pattern for}%
\typeout{** the default language instead.}%
\else
\language=\csname l@#1\endcsname
\fi
#2}}
\providecommand{\BIBdecl}{\relax}
\BIBdecl

\bibitem{Chow2015}
\BIBentryALTinterwordspacing
J.~M. Chow, S.~J. Srinivasan, E.~Magesan, A.~D. Córcoles, D.~W. Abraham, J.~M. Gambetta, and M.~Steffen, ``Characterizing a four-qubit planar lattice for arbitrary error detection,'' in \emph{Quantum Information and Computation XIII}, E.~Donkor, A.~R. Pirich, and M.~Hayduk, Eds., vol. 9500.\hskip 1em plus 0.5em minus 0.4em\relax SPIE, May 2015, p. 95001G. [Online]. Available: \url{http://dx.doi.org/10.1117/12.2192740}
\BIBentrySTDinterwordspacing

\bibitem{Li2023}
\BIBentryALTinterwordspacing
H.-X. Li, D.~Shiri, S.~Kosen, M.~Rommel, L.~Chayanun, A.~Nylander, R.~Rehammar, G.~Tancredi, M.~Caputo, K.~Grigoras, L.~Gr\"{o}nberg, J.~Govenius, and J.~Bylander, ``Experimentally verified, fast analytic, and numerical design of superconducting resonators in flip-chip architectures,'' \emph{IEEE Transactions on Quantum Engineering}, vol.~4, p. 1–12, 2023. [Online]. Available: \url{http://dx.doi.org/10.1109/TQE.2023.3302371}
\BIBentrySTDinterwordspacing

\bibitem{Yen2024}
\BIBentryALTinterwordspacing
A.~Yen, Y.~Ye, K.~Peng, J.~Wang, G.~Cunningham, M.~Gingras, B.~M. Niedzielski, H.~Stickler, K.~Serniak, M.~E. Schwartz, and K.~P. O’Brien, ``Directional emission of a readout resonator for qubit measurement,'' \emph{Physical Review Applied}, vol.~22, no.~3, Sep. 2024. [Online]. Available: \url{http://dx.doi.org/10.1103/PhysRevApplied.22.034035}
\BIBentrySTDinterwordspacing

\bibitem{Sank2024}
\BIBentryALTinterwordspacing
D.~Sank, A.~Opremcak, A.~Bengtsson, M.~Khezri, Z.~Chen, O.~Naaman, and A.~Korotkov, ``System characterization of dispersive readout in superconducting qubits,'' 2024. [Online]. Available: \url{https://arxiv.org/abs/2402.00413}
\BIBentrySTDinterwordspacing

\bibitem{Ma2017}
\BIBentryALTinterwordspacing
P.~Ma, B.~Wei, Y.~Heng, C.~Luo, X.~Guo, and B.~Cao, ``Design of absorptive superconducting filter,'' \emph{Electronics Letters}, vol.~53, no.~11, p. 728–730, May 2017. [Online]. Available: \url{http://dx.doi.org/10.1049/el.2017.0768}
\BIBentrySTDinterwordspacing

\bibitem{Bengtsson2024}
\BIBentryALTinterwordspacing
A.~Bengtsson, A.~Opremcak, M.~Khezri, D.~Sank, A.~Bourassa, K.~J. Satzinger, S.~Hong, C.~Erickson, B.~J. Lester, K.~C. Miao, A.~N. Korotkov, J.~Kelly, Z.~Chen, and P.~V. Klimov, ``Model-based optimization of superconducting qubit readout,'' \emph{Physical Review Letters}, vol. 132, no.~10, Mar. 2024. [Online]. Available: \url{http://dx.doi.org/10.1103/PhysRevLett.132.100603}
\BIBentrySTDinterwordspacing

\bibitem{Peng2022}
\BIBentryALTinterwordspacing
K.~Peng, M.~Naghiloo, J.~Wang, G.~D. Cunningham, Y.~Ye, and K.~P. O’Brien, ``Floquet-mode traveling-wave parametric amplifiers,'' \emph{PRX Quantum}, vol.~3, no.~2, Apr. 2022. [Online]. Available: \url{http://dx.doi.org/10.1103/PRXQuantum.3.020306}
\BIBentrySTDinterwordspacing

\bibitem{Yan2016}
\BIBentryALTinterwordspacing
F.~Yan, S.~Gustavsson, A.~Kamal, J.~Birenbaum, A.~P. Sears, D.~Hover, T.~J. Gudmundsen, D.~Rosenberg, G.~Samach, S.~Weber, J.~L. Yoder, T.~P. Orlando, J.~Clarke, A.~J. Kerman, and W.~D. Oliver, ``The flux qubit revisited to enhance coherence and reproducibility,'' \emph{Nature Communications}, vol.~7, no.~1, Nov. 2016. [Online]. Available: \url{http://dx.doi.org/10.1038/ncomms12964}
\BIBentrySTDinterwordspacing

\bibitem{Gao2021}
\BIBentryALTinterwordspacing
Y.~Y. Gao, M.~A. Rol, S.~Touzard, and C.~Wang, ``Practical guide for building superconducting quantum devices,'' \emph{PRX Quantum}, vol.~2, no.~4, Nov. 2021. [Online]. Available: \url{http://dx.doi.org/10.1103/PRXQuantum.2.040202}
\BIBentrySTDinterwordspacing

\bibitem{Milliken2007}
\BIBentryALTinterwordspacing
F.~P. Milliken, J.~R. Rozen, G.~A. Keefe, and R.~H. Koch, ``50$\omega$ characteristic impedance low-pass metal powder filters,'' \emph{Review of Scientific Instruments}, vol.~78, no.~2, Feb. 2007. [Online]. Available: \url{http://dx.doi.org/10.1063/1.2431770}
\BIBentrySTDinterwordspacing

\bibitem{kow}
\BIBentryALTinterwordspacing
C.~S. Kow and M.~T. Bell, ``Traveling-wave parametric amplifier with passive reverse isolation,'' 2025. [Online]. Available: \url{https://arxiv.org/abs/2505.04059}
\BIBentrySTDinterwordspacing

\bibitem{morgan2011}
M.~A. Morgan and T.~A. Boyd, ``Theoretical and experimental study of a new class of reflectionless filter,'' \emph{IEEE Transactions on Microwave Theory and Techniques}, vol.~59, no.~5, May 2011.

\bibitem{Ge2021}
\BIBentryALTinterwordspacing
Z.~Ge, L.~Chen, L.~Yang, R.~Gomez-Garcia, and X.~Zhu, ``On-chip millimeter-wave integrated absorptive bandstop filter in (bi)-cmos technology,'' \emph{IEEE Electron Device Letters}, vol.~42, no.~1, p. 114–117, Jan. 2021. [Online]. Available: \url{http://dx.doi.org/10.1109/LED.2020.3036036}
\BIBentrySTDinterwordspacing

\bibitem{Richardson2020}
\BIBentryALTinterwordspacing
C.~J.~K. Richardson, A.~Alexander, C.~G. Weddle, B.~Arey, and M.~Olszta, ``Low-loss superconducting titanium nitride grown using plasma-assisted molecular beam epitaxy,'' \emph{Journal of Applied Physics}, vol. 127, no.~23, Jun. 2020. [Online]. Available: \url{http://dx.doi.org/10.1063/5.0008010}
\BIBentrySTDinterwordspacing

\bibitem{obrien}
O'Brien and Peng, ``Josephsoncircuits.jl,'' \url{https://github.com/kpobrien/JosephsonCircuits.jl}, n.d., online; accessed [insert date].

\bibitem{Macklin2015}
\BIBentryALTinterwordspacing
C.~Macklin, K.~O’Brien, D.~Hover, M.~E. Schwartz, V.~Bolkhovsky, X.~Zhang, W.~D. Oliver, and I.~Siddiqi, ``A near–quantum-limited josephson traveling-wave parametric amplifier,'' \emph{Science}, vol. 350, no. 6258, p. 307–310, Oct. 2015. [Online]. Available: \url{http://dx.doi.org/10.1126/science.aaa8525}
\BIBentrySTDinterwordspacing

\bibitem{Kurokawa1965}
\BIBentryALTinterwordspacing
K.~Kurokawa, ``Power waves and the scattering matrix,'' \emph{IEEE Transactions on Microwave Theory and Techniques}, vol.~13, no.~2, p. 194–202, Mar. 1965. [Online]. Available: \url{http://dx.doi.org/10.1109/TMTT.1965.1125964}
\BIBentrySTDinterwordspacing

\bibitem{zverev1967handbook}
A.~I. Zverev, \emph{Handbook of Filter Synthesis}.\hskip 1em plus 0.5em minus 0.4em\relax Wiley-Interscience, 1967.

\bibitem{Haus1987}
\BIBentryALTinterwordspacing
H.~Haus, W.~Huang, S.~Kawakami, and N.~Whitaker, ``Coupled-mode theory of optical waveguides,'' \emph{Journal of Lightwave Technology}, vol.~5, no.~1, p. 16–23, 1987. [Online]. Available: \url{http://dx.doi.org/10.1109/JLT.1987.1075416}
\BIBentrySTDinterwordspacing

\bibitem{reed1956}
J.~Reed and G.~J. Wheeler, ``A method of analysis of symmetrical four-port networks,'' \emph{IEEE Transactions on Microwave Theory and Techniques}, 1956.

\bibitem{zhang2019}
F.~Zhang, C.~Ma, and T.~Zhao, ``Modeling of mutual inductance between planar inductors on the same plane,'' in \emph{Proceedings of the 2019 8th International Symposium on Next Generation Electronics (ISNE)}, 2019.

\bibitem{Clerk2010}
\BIBentryALTinterwordspacing
A.~A. Clerk, M.~H. Devoret, S.~M. Girvin, F.~Marquardt, and R.~J. Schoelkopf, ``Introduction to quantum noise, measurement, and amplification,'' \emph{Reviews of Modern Physics}, vol.~82, no.~2, p. 1155–1208, Apr. 2010. [Online]. Available: \url{http://dx.doi.org/10.1103/RevModPhys.82.1155}
\BIBentrySTDinterwordspacing

\bibitem{lebl2019}
\BIBentryALTinterwordspacing
A.~Lebl, M.~Mileusni, B.~Pavic, and J.~Radivojevic, ``Absorptive filters in the realization of rcied activation jamming,'' in \emph{2019 IEEE 31st International Conference on Microelectronics (MIEL)}.\hskip 1em plus 0.5em minus 0.4em\relax IEEE, Sep. 2019, p. 231–234. [Online]. Available: \url{http://dx.doi.org/10.1109/MIEL.2019.8889582}
\BIBentrySTDinterwordspacing

\bibitem{neumann1846}
F.~E. Neumann, ``Allgemeine gesetze der inducierten elektrischen ströme,'' \emph{Annalen der Physik}, vol. 143, pp. 31--44, 1846.

\bibitem{yeh2017}
J.-H. Yeh, J.~LeFebvre, S.~Premaratne, F.~C. Wellstood, and B.~S. Palmer, ``Microwave attenuators for use with quantum devices below 100 mk,'' \emph{Journal of Applied Physics}, vol. 121, no.~22, p. 224501, 2017.

\bibitem{scikit}
A.~Arsenovic, J.~Hillairet, J.~Anderson, H.~Forstén, V.~Rieß, M.~Eller, N.~Sauber, R.~Weikle, W.~Barnhart, and F.~Forstmayr, ``scikit-rf: An open source python package for microwave network creation, analysis, and calibration [speaker’s corner],'' \emph{IEEE Microwave Magazine}, vol.~23, no.~1, pp. 98--105, 2022.

\bibitem{Dunsworth2018}
\BIBentryALTinterwordspacing
A.~Dunsworth, R.~Barends, Y.~Chen, Z.~Chen, B.~Chiaro, A.~Fowler, B.~Foxen, E.~Jeffrey, J.~Kelly, P.~Klimov, E.~Lucero, J.~Mutus, M.~Neeley, C.~Neill, C.~Quintana, P.~Roushan, D.~Sank, A.~Vainsencher, J.~Wenner, T.~White, H.~Neven, J.~Martinis, and A.~Megrant, ``A method for building low loss multi-layer wiring for superconducting microwave devices,'' \emph{Applied Physics Letters}, vol. 112, no.~6, Mar. 2018. [Online]. Available: \url{https://doi.org/10.1063/1.5014033}
\BIBentrySTDinterwordspacing

\bibitem{Rome1973-nichrome}
\BIBentryALTinterwordspacing
{Rome Air Development Center}, ``Nichrome resistor properties and reliability,'' Rome Air Development Center, Rome, NY, Tech. Rep. {AD-765 534}, Jun. 1973. [Online]. Available: \url{https://apps.dtic.mil/sti/tr/pdf/AD0765534.pdf}
\BIBentrySTDinterwordspacing

\bibitem{Wang2021-hi}
H.~Wang, S.~Singh, C.~R.~H. McRae, J.~C. Bardin, S.-X. Lin, N.~Messaoudi, A.~R. Castelli, Y.~J. Rosen, E.~T. Holland, D.~P. Pappas, and J.~Y. Mutus, ``Cryogenic single-port calibration for superconducting microwave resonator measurements,'' \emph{Quantum Sci. Technol.}, vol.~6, no.~3, p. 035015, Jul. 2021.

\bibitem{rytting}
D.~Rytting, ``Network analyzer error models and calibration methods,'' blub, Tech. Rep., n.d., online; \url{http://emlab.uiuc.edu/ece451/appnotes/Rytting_NAModels.pdf}.

\bibitem{Rehnmark1974}
\BIBentryALTinterwordspacing
S.~Rehnmark, ``On the calibration process of automatic network analyzer systems (short papers),'' \emph{IEEE Transactions on Microwave Theory and Techniques}, vol.~22, no.~4, p. 457–458, Apr. 1974. [Online]. Available: \url{http://dx.doi.org/10.1109/TMTT.1974.1128250}
\BIBentrySTDinterwordspacing

\bibitem{Glasser1978}
\BIBentryALTinterwordspacing
L.~Glasser, ``An analysis of microwave de-embedding errors (technical notes),'' \emph{IEEE Transactions on Microwave Theory and Techniques}, vol.~26, no.~5, p. 379–380, May 1978. [Online]. Available: \url{http://dx.doi.org/10.1109/TMTT.1978.1129395}
\BIBentrySTDinterwordspacing

\bibitem{Li2021}
\BIBentryALTinterwordspacing
X.~Li, M.~Xing, G.~Liu, X.~Yang, C.~Dai, and M.~Hou, ``Compact, reflectionless band-pass filter: Based on gaas ipd process for highly reliable communication,'' \emph{Electronics}, vol.~10, no.~23, p. 2998, Dec. 2021. [Online]. Available: \url{https://doi.org/10.3390/electronics10232998}
\BIBentrySTDinterwordspacing

\bibitem{Simpson2021}
D.~Simpson and D.~Psychogiou, ``X-band quasi-reflectionless mmic bandpass filters with minimum number of components,'' \emph{IEEE Transactions on Electron Devices}, vol.~68, no.~9, pp. 4329--4334, jul 2021.

\bibitem{Liu2020}
G.~Liu, M.~Xing, X.~Li, S.~Xu, and C.~Dai, ``Design of a miniaturized reflectionless bandpass filter with high selectivity for 5g network,'' in \emph{Recent Developments in Mechatronics and Intelligent Robotics (ICMIR 2019)}, ser. Advances in Intelligent Systems and Computing, S.~Patnaik, J.~Wang, Z.~Yu, and N.~Dey, Eds.\hskip 1em plus 0.5em minus 0.4em\relax Springer, Singapore, 2020, vol. 1060, pp. 585--593, first Online: 05 March 2020.

\bibitem{minicircuitsReflectionlessPass}
``{R}eflectionless {L}ow {P}ass {F}ilter, {D}{C} - 6000 {M}{H}z, 50\&x3{A}9; | {X}{L}{F}-63+ | {M}ini-{C}ircuits --- minicircuits.com,'' \url{https://www.minicircuits.com/WebStore/dashboard.html?model=XLF-63%2B&srsltid=AfmBOooeqVsJ1ZH8hqfOylUA4k11gI43f2lnBPJ07i0MSag5n-IBwNQ7}, [Accessed 08-04-2025].

\bibitem{Simbierowicz2024}
\BIBentryALTinterwordspacing
S.~Simbierowicz, M.~Borrelli, V.~Monarkha, V.~Nuutinen, and R.~E. Lake, ``Inherent thermal-noise problem in addressing qubits,'' \emph{PRX Quantum}, vol.~5, no.~3, Jul. 2024. [Online]. Available: \url{http://dx.doi.org/10.1103/PRXQuantum.5.030302}
\BIBentrySTDinterwordspacing

\bibitem{Guarcello2024}
\BIBentryALTinterwordspacing
C.~Guarcello, C.~Barone, G.~Carapella, V.~Granata, G.~Filatrella, A.~Giachero, and S.~Pagano, ``Driving a josephson traveling wave parametric amplifier into chaos: Effects of a non-sinusoidal current–phase relation,'' \emph{Chaos, Solitons \& Fractals}, vol. 189, p. 115598, Dec. 2024. [Online]. Available: \url{http://dx.doi.org/10.1016/j.chaos.2024.115598}
\BIBentrySTDinterwordspacing

\bibitem{Faramarzi24}
\BIBentryALTinterwordspacing
F.~Faramarzi, R.~Stephenson, S.~Sypkens, B.~H. Eom, H.~LeDuc, and P.~Day, ``A 4-8 ghz kinetic inductance travelling-wave parametric amplifier using four-wave mixing with near quantum-limit noise performance,'' 2024. [Online]. Available: \url{https://arxiv.org/abs/2402.11751}
\BIBentrySTDinterwordspacing

\bibitem{Gaydamachenko25}
\BIBentryALTinterwordspacing
V.~Gaydamachenko, C.~Kissling, and L.~Gr\"{u}nhaupt, ``An rf-squid-based traveling-wave parametric amplifier with -84 dbm input saturation power across more than one octave bandwidth,'' 2025. [Online]. Available: \url{https://arxiv.org/abs/2503.02489}
\BIBentrySTDinterwordspacing

\bibitem{Gunyh2024}
\BIBentryALTinterwordspacing
A.~M. Gunyhó, S.~Kundu, J.~Ma, W.~Liu, S.~Niemel\"{a}, G.~Catto, V.~Vadimov, V.~Vesterinen, P.~Singh, Q.~Chen, and M.~M\"{o}tt\"{o}nen, ``Single-shot readout of a superconducting qubit using a thermal detector,'' \emph{Nature Electronics}, vol.~7, no.~4, p. 288–298, Apr. 2024. [Online]. Available: \url{http://dx.doi.org/10.1038/s41928-024-01147-7}
\BIBentrySTDinterwordspacing

\bibitem{Sah2024}
\BIBentryALTinterwordspacing
A.~Sah, S.~Kundu, H.~Suominen, Q.~Chen, and M.~M\"{o}tt\"{o}nen, ``Decay-protected superconducting qubit with fast control enabled by integrated on-chip filters,'' \emph{Communications Physics}, vol.~7, no.~1, Jul. 2024. [Online]. Available: \url{http://dx.doi.org/10.1038/s42005-024-01733-3}
\BIBentrySTDinterwordspacing

\bibitem{Krantz2019}
\BIBentryALTinterwordspacing
P.~Krantz, M.~Kjaergaard, F.~Yan, T.~P. Orlando, S.~Gustavsson, and W.~D. Oliver, ``A quantum engineer’s guide to superconducting qubits,'' \emph{Applied Physics Reviews}, vol.~6, no.~2, Jun. 2019. [Online]. Available: \url{http://dx.doi.org/10.1063/1.5089550}
\BIBentrySTDinterwordspacing

\end{thebibliography}

\vspace{-30pt} %
\begin{IEEEbiographynophoto}{Jessica Kedziora}
received the S.M. degree from the University of California, San Diego, in 2018. She is currently pursuing a Ph.D. in Electrical Engineering and Computer Science at the Massachusetts Institute of Technology. She has over 30 years of experience in the aerospace and defense industry, including 13 years running a consulting firm specializing in integrated microwave assemblies for RADAR, sensing, guidance, and military communications. She is a member of the IEEE and the American Physical Society.
\end{IEEEbiographynophoto}
\vspace{-20pt}
\begin{IEEEbiographynophoto}{Eric Bui}
received his B.S. degree in Electrical Engineering and Computer Science at Massachusetts Institute of Technology in 2024. He is currently enrolled in the EECS masters program at MIT and will be joining as a PhD student in the fall of 2025. He spent three years at the MIT.nano cleanroom facility, developing processes for superconducting electronics. 
\end{IEEEbiographynophoto}


\end{document}